\documentclass[fleqn,usenatbib]{mnras}
\pdfoutput=1
\usepackage{newtxtext,newtxmath}

\usepackage[T1]{fontenc}

\DeclareRobustCommand{\VAN}[3]{#2}
\let\VANthebibliography\thebibliography
\def\thebibliography{\DeclareRobustCommand{\VAN}[3]{##3}\VANthebibliography}
\usepackage[normalem]{ulem}

\usepackage{graphicx}	
\usepackage{amsmath}	
\usepackage{mathtools}

\title[Residual H/He atmospheres of super-Earths]{To cool is to keep: Residual H/He atmospheres of super-Earths and sub-Neptunes}

\author[W. Misener and H.E. Schlichting]{
William Misener$^{1}$
and Hilke E. Schlichting$^{1,2}$
\\
$^{1}$Department of Earth, Planetary, and Space Sciences, The University of California, Los Angeles, 595 Charles E. Young Drive East, Los Angeles, CA 90095, USA\\
$^{2}$Department of Earth, Atmospheric and Planetary Sciences, Massachusetts Institute of Technology, 77 Massachusetts Avenue, Cambridge, MA 02139, USA}

\date{Accepted XXX. Received YYY; in original form ZZZ}

\pubyear{2021}

\begin{document}
\label{firstpage}
\pagerange{\pageref{firstpage}--\pageref{lastpage}}
\maketitle

\begin{abstract}
Super-Earths and sub-Neptunes are commonly thought to have accreted hydrogen/helium envelopes, consisting of a few to ten percent of their total mass, from the primordial gas disk. Subsequently, hydrodynamic escape driven by core-powered mass-loss and/or photo-evaporation likely stripped much of these primordial envelopes from the lower-mass and closer-in planets to form the super-Earth population. In this work we show that after undergoing core-powered mass-loss, some super-Earths can retain small residual H/He envelopes. This retention is possible because, for significantly depleted atmospheres, the density at the radiative-convective boundary drops sufficiently such that the cooling time-scale becomes shorter than the mass-loss time-scale. The residual envelope is therefore able to contract, terminating further mass loss. Using analytic calculations and numerical simulations, we show that the mass of primordial H/He envelope retained as a fraction of the planet’s total mass, $f_\mathrm{ret}$, increases with increasing planet mass, $M_\mathrm{c}$, and decreases with increasing equilibrium temperature, $T_\mathrm{eq}$, scaling as $f_\mathrm{ret} \propto M_\mathrm{c}^{3/2} T_\mathrm{eq}^{-1/2} \exp{[M_\mathrm{c}^{3/4} T_\mathrm{eq}^{-1}]}$. $f_\mathrm{ret}$ varies from $<10^{-8}$ to about $10^{-3}$ for typical super-Earth parameters. To first order, the exact amount of left-over H/He depends on the initial envelope mass, the planet mass, its equilibrium temperature, and the envelope's opacity. These residual hydrogen envelopes reduce the atmosphere's mean molecular weight compared to a purely secondary atmosphere, a signature observable by current and future facilities. These remnant atmospheres may, however, in many cases be vulnerable to long-term erosion by photo-evaporation. Any residual hydrogen envelope likely plays an important role in the long-term physical evolution  of super-Earths, including their geology and geochemistry. 
\end{abstract}




\section{Introduction}

Close-in exoplanets with sizes between that of Earth and Neptune are the most common exoplanets in our galaxy known to date \citep[e.g.][]{PHM13, F13}. Significant work has been dedicated to understanding their formation and subsequent evolution \citep[e.g.][]{HM12, S14, IS15, LC15, GSS16, IO17}. The radii of a significant fraction of these planets are sufficiently large that they show evidence for substantial H/He envelopes containing a few percent of the planets' total mass \citep[e.g.][]{WL15}. This implies that these worlds formed in the presence of the primordial gas disk. Since atmospheric mass loss is common after the dispersal of the gas disk, atmospheric masses that are observed today are generally not the same as those accreted in the presence of the primordial gas disk \citep[e.g.][]{IH12, OJ12, LFM12, OW16, GSS16}. As a result, even super-Earths that appear as barren rocky cores today are consistent with having formed with primordial H/He envelopes \citep[e.g.][]{S18}. The two prevailing mechanisms which aim to explain the loss of these envelopes are photo-evaporation due to high energy flux from the host star \citep[e.g.][]{OJ12,LFM12} and core-powered mass loss, during which the cooling luminosity from the hot underlying planetary core fuels the atmospheric loss \citep[e.g.][]{GSS16}. Both of these mechanisms yield a double peaked radius distribution of exoplanet sizes \citep[e.g.][]{OW17, GS19, GS20} consistent with observations \citep[e.g.][]{OW13, FP17, vE18}. While recent observations determining the ages of planet-hosting stars report a significant increase in super-Earths relative to sub-Neptunes on a gigayear time-scale, confirming one of the predictions from core-powered mass loss models \citep{B20,D20}, there is not yet consensus as to which mechanism dominates, if any, in sculpting the observed bimodality \citep[e.g.][]{LS20}.

\begin{figure*}
\centering
\includegraphics{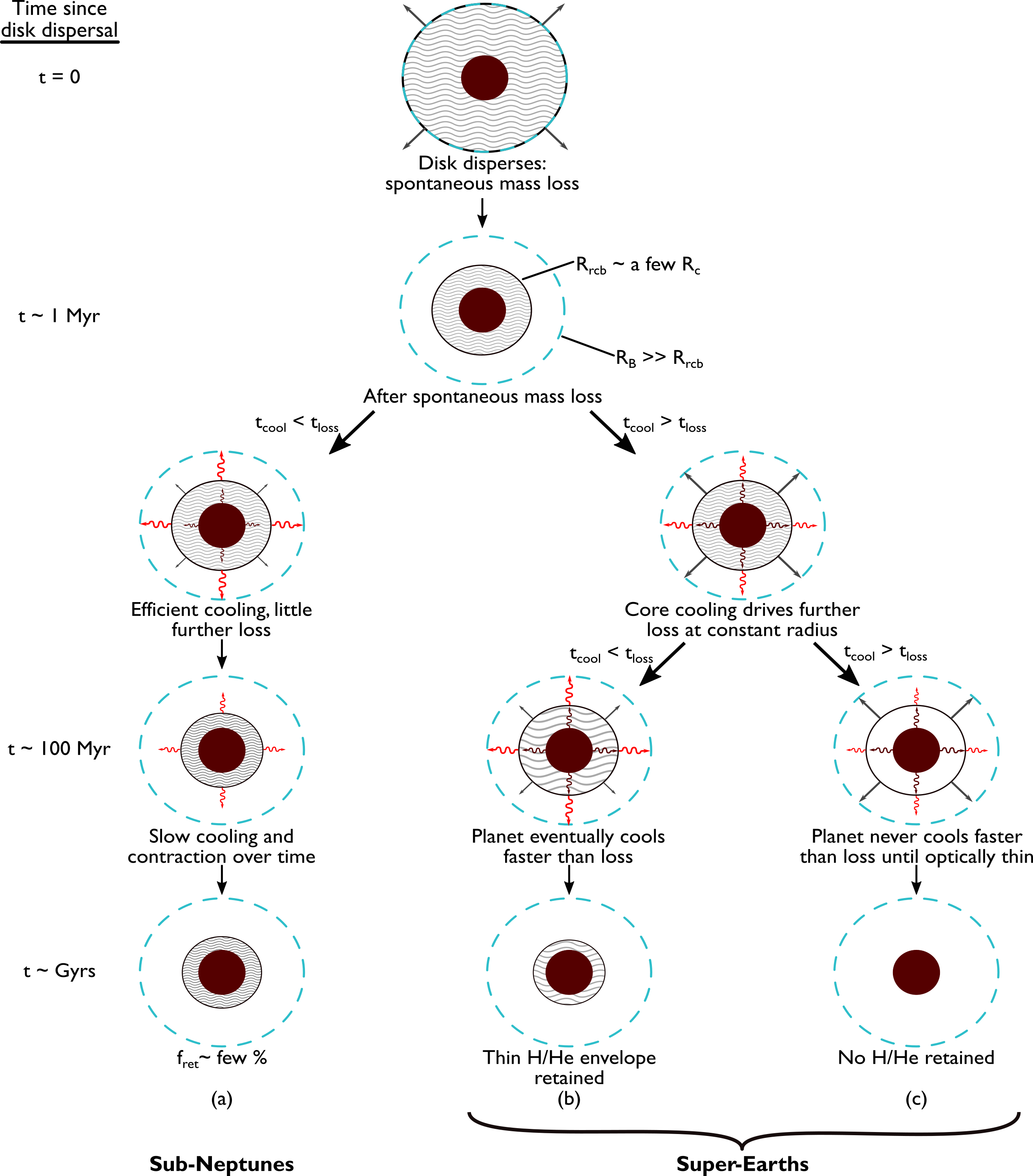}
\caption{A schematic of sub-Neptune and super-Earth formation and evolution. Thermal cooling is shown by wavy arrows, while mass loss is depicted by straight gray arrows, where only the relative sizes of the arrows are important. A planet begins as a core (in maroon) embedded in the gaseous protoplanetary disk. It accretes an H/He envelope (in gray) which is bound interior to its Bondi radius, $R_\mathrm{B}$ (blue dashed circle). When the disk disperses, the planet quickly sheds its outer layers and contracts. Its radiative-convective boundary, $R_\mathrm{rcb}$, (black circle) shrinks to a few core radii, $R_\mathrm{c}$. In Scenario (a), the planet's available energy for cooling is dominated by the atmosphere at the end of spontaneous mass loss phase, so it can efficiently cool and contract, cutting off any further loss. The planet thus retains much of its primordial envelope and becomes a sub-Neptune with $f_\mathrm{ret}$ of order a few percent. If the thermal energy stored in the core instead dominates the available energy budget for cooling, the thermal energy released by the core into the atmosphere keeps the atmosphere inflated at a nearly constant $R_\mathrm{rcb}$, which ensures continued atmospheric loss. Mass loss continues until either the atmosphere has lost so much mass that its density at $R_\mathrm{rcb}$ decreases sufficiently that it can start to cool more quickly than mass is lost, preserving some primordial H/He as in Scenario (b), or until the atmosphere becomes optically thin to outgoing radiation after nearly all H/He is lost, allowing the core to cool directly to space, illustrated by Scenario (c).}
\label{fig:diagram}
\end{figure*}

Figure \ref{fig:diagram} schematically summarizes the key steps in the atmospheric mass-loss histories of super-Earths and sub-Neptunes. First, once the disk disperses, recently accreted envelopes are partially lost due to the loss of pressure support from the surrounding disk, provided the disk dispersal time-scale is shorter than the envelopes' cooling time-scales. As a result, planets shed their outer layers, a process termed spontaneous mass loss \citep[e.g.][]{IH12, OW16, GSS16}. In the spontaneous mass loss phase, planets will lose several tens of percent of their envelope masses and shrink to envelope thicknesses that are of order their core radii on roughly megayear time-scales.

In the photo-evaporation model, atmospheric loss following this phase is powered by high energy radiation from the host star. If the time-scale to evaporate the atmosphere remains shorter than the duration of increased high energy flux from the star, the planet continues to lose mass until a bare super-Earth core remains. Otherwise, if insufficient energy is received to unbind the atmosphere, the planet retains most of its primordial gaseous envelope, roughly doubling its observed radius and corresponding to the observed sub-Neptune population. This loss process has been demonstrated to be consistent with the observed bimodal exoplanet radius distribution \citep[e.g.][]{OJ12, LFM12, OW17}.

In contrast, in the core-powered mass loss model, the future of the planet's envelope is determined by its own thermal energy available for cooling and the bolometric luminosity of the host star, as this sets the gasses' escape speed at the sonic radius. If the core's heat capacity is negligible compared to that of the atmosphere after spontaneous mass loss, there is insufficient energy for loss and the atmosphere will remain bound. The planet will continue to cool and contract, which will quickly halt any further mass loss. This process leads to a planet which retains a substantial H/He envelope of order a few percent of the planet's total mass, corresponding to the observed sub-Neptune population. This general evolution is shown in Scenario (a) of Figure \ref{fig:diagram}. Conversely, if the core has a larger heat capacity than the remaining atmosphere, the atmosphere's thermal evolution is coupled to that of the core. The core is in thermal equilibrium with the base of the atmosphere, so as the upper atmosphere cools radiatively to space, the core also cools and resupplies heat to the atmosphere. This transfer of energy from the core to the atmosphere prevents further atmospheric contraction, keeping the atmosphere inflated and thus driving further mass loss. In this way the planet can lose much of its remaining envelope and become a super-Earth, as shown in previous work \citep{GSS16, GS19, GS20}.

However, unlike in photo-evaporation, this atmospheric loss from super-Earths does not necessarily proceed to completion. The atmosphere's luminosity is determined by radiative diffusion across the radiative-convective boundary, $R_\mathrm{rcb}$. As the atmosphere loses mass at constant radius, its density and therefore its optical depth at $R_\mathrm{rcb}$ decreases. Radiative diffusion thus becomes more efficient, and the planet becomes more luminous, decreasing the cooling time-scale. Therefore, just as for the sub-Neptunes, a super-Earth's cooling time-scale can become shorter than its mass loss time-scale. At this point, the atmosphere can once again contract, thereby exponentially increasing the mass loss time-scale and quenching loss. In this fashion, some primordial H/He can be saved by cooling even for super-Earths, as shown by Scenario (b) of Figure \ref{fig:diagram}. However, in cases where mass loss is sufficiently fast, the atmospheric cooling time-scale may always be longer than the mass loss time-scale, and loss proceeds until the atmosphere becomes fully optically thin at very low residual atmospheric mass. For this end-member, virtually all the primordial H/He is lost, and any atmosphere at late times is essentially entirely outgassed (Figure \ref{fig:diagram}, Scenario (c)).

The focus of this work is to investigate what ultimately determines the final primordial H/He atmospheres of close-in exoplanets and how much of their natal envelopes super-Earths and sub-Neptunes can retain after core-powered atmospheric loss. The properties of the residual primordial atmospheres of super-Earths are especially interesting for several reasons. For one, the residual atmospheres set the conditions for outgassing of any secondary atmospheres \citep[e.g.][]{GS14, K20}. In addition, any residual nebular hydrogen will significantly alter the redox state of rocky exoplanets and hence their geology and geochemistry \citep[e.g.][]{WSF18, DY19}. Preliminary investigations of the habitability of hydrogen-dominated super-Earth atmospheres \citep[e.g.][]{S20} will gain more importance if such conditions are expected to be common. Finally, if rocky super-Earths can retain some residual hydrogen, their final envelopes will have lower mean molecular weights than pure secondary atmospheres, and correspondingly larger scale heights. These H-rich atmospheres could be distinguished from pure secondary atmospheres composed of heavier species by observations of their transmission spectra, features of which are sensitive to mean molecular weight \citep[e.g.][]{BS12, FM13}. Such observational tests are already possible for sub-Neptunes \citep[e.g.][]{BW19}, and upcoming facilities such as the \textit{James Webb Space Telescope} and \textit{Ariel} should also be able to distinguish H-rich super-Earth atmospheres from those with higher mean molecular weights \citep[e.g.][]{GL16, E19}.

In the following sections, we calculate the amount of primordial H/He that a super-Earth is expected to retain assuming its evolution is dominated by core-powered mass-loss. We quantify the different stages of evolution through which these planets progress, and we show that our predictions will be testable by future observations. The paper is structured as follows. We define the key concepts of our mass-loss and thermal evolution models in Section \ref{sec:basics}. In Section \ref{sec:analytic} we determine the mass-loss and cooling time-scales in two regimes: the `spontaneous mass loss' regime in which loss is driven by the atmosphere's own energy, and the `core-powered' regime in which loss is instead driven by the cooling luminosity of the core. We describe how core-powered mass loss can naturally cease at low atmospheric masses and analytically determine the residual H/He envelope masses as functions of the planet mass and equilibrium temperature. To complement our analytic approximations, we present a numerical model of these phases of atmospheric evolution in Section \ref{sec:numerical} and compare the results to our analytic expressions. In Section \ref{sec:obs} we present observational predictions. Discussions and conclusions follow in Sections \ref{sec:discussion} and \ref{sec:conclusions}, respectively.

\section{Model and Approach}\label{sec:basics}
In this section, we summarize our model. We describe the key physical parameters that govern the mass loss and cooling of super-Earths, including our model of the core, the atmosphere, and their evolution over time.

\subsection{Core model}
We assume planets form in the protoplanetary disk as rocky cores of mass $M_\mathrm{c} \sim$ a few $M_\oplus$. We take the cores as similar to Earth in bulk density, accounting for compression. Their radii, $R_\mathrm{c}$, are thus determined by the mass-radius relation appropriate for rocky cores: $R_\mathrm{c}/R_\oplus = (M_\mathrm{c}/M_\oplus)^{1/\beta}$ where $R_\oplus$ and $M_\oplus$ are the radius and mass of Earth, respectively, and $\beta \simeq 4$ \citep[e.g.][]{V06,S07}. These assumptions are consistent with the underlying planet properties derived from both photo-evaporation \citep{RO20} and core-powered mass-loss models \citep{GS19}. The planet's atmosphere is typically on the order of a few percent or less for the evolution we consider here, and so we ignore the atmosphere's effect on the planet's overall gravity.

We model the core as incompressible and isothermal, with a temperature, $T_\mathrm{c}$, that is coupled to the atmospheric temperature at its surface. For the purposes of this analysis, incorporating the true thermal gradient inside the core would not significantly alter the results. Therefore the thermal energy in the core is approximately
\begin{equation}\label{eq:energycore}
    E_\mathrm{c} = C_\mathrm{c} T_\mathrm{c} \simeq \frac{1}{\gamma_\mathrm{c}-1} N k_\mathrm{B} T_\mathrm{c}\mathrm{,}
\end{equation}
where $C_\mathrm{c}$ is the heat capacity of the core. In the second equality, we model the heat capacity using the form for an ideal gas, where
$\gamma_\mathrm{c}$ is the adiabatic index of the core. Throughout this work, we use $\gamma_\mathrm{c} = 4/3$, following the Dulong-Petit law. This is an upper limit on the core's effective adiabatic index, as liquid silicates can have heat capacities higher than those of solids \citep[e.g.][]{SSD17}. The number of molecules in the core is given by $N = M_\mathrm{c}/\mu_\mathrm{c}$, where $\mu_\mathrm{c}$ is the mean molecular weight of the core. Motivated by our own Earth, we assume $\mu_\mathrm{c}= 60$ amu.

\subsection{Atmospheric structure}
While embedded in the primordial disk, a planet will capture any gas within its sphere of influence, which we approximate by the Bondi radius, $R_\mathrm{B}$. The Bondi radius is the point where the escape velocity from the planet is equal to the thermal velocity of the gas molecules. The isothermal speed of sound is given by $c_\mathrm{s} = (k_\mathrm{B} T/\mu)^{1/2}$, where $k_\mathrm{B}$ is the Boltzmann constant, $T$ is the temperature at $R_\mathrm{B}$, and $\mu$ the mean molecular weight of the atmosphere. We assume throughout that $\mu = 2.2$ amu, as expected for a primordial H/He atmosphere. We can write the Bondi radius as $R_\mathrm{B} \simeq 2 G M_\mathrm{c} \mu/(k_\mathrm{B} T_\mathrm{eq})$, where $G$ is the gravitational constant and $T_\mathrm{eq}$ is the equilibrium temperature, determined by the incident flux of stellar radiation. The equilibrium temperature scales with the planet's semi-major axis, $a$, as
\begin{equation}\label{eq:tempa}
    T_\mathrm{eq} = \bigg(\frac{L_*}{16 \pi \sigma a^2}\bigg)^{1/4} = 279\ \mathrm{K} \bigg(\frac{a}{1\ \mathrm{au}}\bigg)^{-1/2}\mathrm{,}
\end{equation}
where $L_*$ is the luminosity of the planet's host star, $\sigma$ is the Stefan-Boltzmann constant, and the right-most expression is evaluated using $L_*=L_\odot$, matching the classic \citet{H81} profile. We paramaterize our results throughout this work in terms of $T_\mathrm{eq}$, as this temperature determines the physical outflow behavior.

We model the structure of the planetary atmosphere as an inner convective region with an adiabatic profile and an outer radiative region that is isothermal at the equilibrium temperature $T_\mathrm{eq}$ \citep[e.g.][]{LC15,GSS16}. The transition between the two regions is labeled the radiative-convective boundary, $R_\mathrm{rcb}$. Thus the density structure of the inner region $R_\mathrm{c} \leq r \leq R_\mathrm{rcb}$ is
\begin{equation}\label{eq:density_struc}
    \rho(r) = \rho_\mathrm{rcb} \bigg(1+\frac{R_\mathrm{B}'}{r} - \frac{R_\mathrm{B}'}{R_\mathrm{rcb}}\bigg)^{1/(\gamma - 1)}\mathrm{,}
\end{equation}
where $r$ and $R_\mathrm{rcb}$ are measured from the center of the planet, $\gamma$ is the adiabatic index of the atmosphere, $R_\mathrm{B}' \equiv (\gamma-1)/(2 \gamma) \times R_\mathrm{B}$ is defined for convenience, and $\rho_\mathrm{rcb} = \rho(R_\mathrm{rcb})$. The density structure in the outer region, $R_\mathrm{rcb} \leq r \leq R_\mathrm{B}$, is well-described by
\begin{equation}\label{eq:density_iso}
    \rho(r) \simeq  \rho_\mathrm{rcb} \exp{\bigg[\frac{R_\mathrm{B}}{2r} -\frac{R_\mathrm{B}} {2R_\mathrm{rcb}} \bigg]} \mathrm{.}
\end{equation}
Since the density decays exponentially past $R_\mathrm{rcb}$, almost all the atmospheric mass is contained in the inner convective region, so we approximate the atmospheric mass as the integral of the adiabatic density profile
\begin{equation}\label{eq:mass_int}
    M_\mathrm{atm} \equiv f M_\mathrm{c} \simeq 4 \pi \rho_\mathrm{rcb} \int_{R_\mathrm{c}}^{R_\mathrm{rcb}} r^2 \bigg(1+\frac{R_\mathrm{B}'}{r} - \frac{R_\mathrm{B}'}{R_\mathrm{rcb}}\bigg)^{1/(\gamma - 1)} dr \mathrm{,}
\end{equation}
where we express the atmospheric mass throughout this work as a fraction of the core's mass, $f \equiv M_\mathrm{atm} /M_\mathrm{c}$, for convenience. We verify \textit{a posteriori} that for the duration of mass loss, less than 10 percent of the total mass of the atmosphere is in the isothermal region for the vast majority of planets considered here. The assumption that we can neglect, to first order, the atmospheric mass in the isothermal region when calculating the residual envelope mass at the point when mass-loss ceases is therefore justified. The only planets for which we found isothermal regions comprising greater than 10 percent of the atmosphere's mass cease losing mass because the atmosphere becomes optically thin (see Section \ref{sec:optthin}), which are not the topic of this paper. The phase space that is relevant to this optically-thin regime is marked as the shaded gray region in Figure \ref{fig:f_contours}.

The atmospheric specific energy is the sum of the (negative) gravitational and (positive) thermal energy and is given by
\begin{equation}\label{eq:energy_specific}
    e(r) = -\frac{G M_\mathrm{c}}{r} + \frac{1}{\gamma-1} \frac{k_\mathrm{B} T(r)}{\mu} \mathrm{,}
\end{equation}
where $T(r)/T_\mathrm{eq} = (\rho/\rho_\mathrm{rcb})^{\gamma-1}$ in the convective region and $T(r \geq R_\mathrm{rcb}) = T_\mathrm{eq}$ in the isothermal region. The total energy, too, is concentrated in the convective region and hence well approximated as
\begin{equation}\label{eq:energy_int}
    E_\mathrm{atm} = \int_{M_\mathrm{atm}} e dm \simeq \int_{R_\mathrm{c}}^{R_\mathrm{rcb}} 4 \pi r^2 e(r) \rho(r) dr \mathrm{.}
\end{equation}

\subsection{Atmospheric evolution}
We fundamentally quantify the evolution of these atmospheres using the mass loss and cooling time-scales, given by $t_\mathrm{loss} = |M_\mathrm{atm}/\dot{M}_\mathrm{atm}|$ and $t_\mathrm{cool} = |E/\dot{E}|$, respectively. If $t_\mathrm{loss} > t_\mathrm{cool}$, the planet can efficiently cool and contract, suppressing any further mass loss. Conversely, if $t_\mathrm{loss} < t_\mathrm{cool}$, hydrodynamic mass loss is more efficient than cooling.

We treat the mass of the planet's H/He envelope at point of disk dispersal, $M_\mathrm{atm, init}$, as an input variable to our model, but assume it is on the order of a few to ten percent of the core's mass \citep[e.g.][]{GSS16}. Motivated by observations \citep[e.g.][]{K13}, we assume that the disk dispersal time-scale is short compared to the cooling time-scale of the envelope. Therefore, after outside pressure support is lost due to disk dispersal, the atmospheres of these planets will quickly lose mass hydrodynamically. The mass loss rate is limited by the rate at which the gas molecules can escape at the outer radius (i.e. the smaller of the Hill and Bondi radius)
\begin{equation}\label{eq:bondiloss}
    \dot{M}_\mathrm{B} = - 4 \pi R_\mathrm{out}^2 u_\mathrm{out} \rho_\mathrm{out} \simeq - 4 \pi R_\mathrm{s}^2 c_\mathrm{s} \rho_\mathrm{rcb} \exp{\bigg[- \frac{2R_\mathrm{s}} {R_\mathrm{rcb}} \bigg]} \mathrm{,}
\end{equation}
where $\rho_\mathrm{out}$ and $u_\mathrm{out}$ are the density and flow speed at the outer radius $R_\mathrm{out}$. In the second equality we have assumed that is outer radius is given by the radius of the sonic point $R_\mathrm{s}$, which we assume for simplicity throughout this paper, rather than the Hill radius.

After disk dispersal, energy is radiated from the planet with a luminosity given by \citep[][Eq 15]{GSS16}
\begin{equation}\label{eq:luminosity}
    L = -\dot{E} \simeq \frac{64\pi \sigma T_\mathrm{eq}^4 R_\mathrm{B}'}{3 \kappa \rho_\mathrm{rcb}} \mathrm{,}
\end{equation}
where $\sigma$ is the Stefan-Boltzmann constant and $\kappa$ the opacity at the $R_\mathrm{rcb}$. We assume a Rosseland mean opacity of the gas with $\kappa=0.1$ cm$^2$g$^{-1}$, a constant suitable for H/He dominated atmospheres \citep[e.g.][]{FML08}. Using a more realistic scaling of opacity with atmospheric density only marginally affects our results, though we discuss other potential scalings in Section \ref{sec:discussion}.
Some portion of this luminosity goes into liberating mass out of the planet's potential. If all of the internal luminosity of the atmosphere goes into mass liberation, we find the maximum rate at which mass can be lost energetically, which we term the luminosity-limited mass loss rate:
\begin{equation}
    \dot{M}_\mathrm{L} \simeq \frac{L M_\mathrm{atm}}{E_\mathrm{atm}}\mathrm{.}
\end{equation}
If $\dot{M}_\mathrm{L} < \dot{M}_\mathrm{B}$, there is insufficient energy available to lift mass to the radiative-convective boundary and maintain the Parker wind-type loss, and mass loss is limited by the underlying luminosity rather than by the rate at which gas molecules can escape from the Bondi radius. In this regime, the planet loses mass and cools on similar time-scales and $t_\mathrm{cool} 
\simeq t_\mathrm{loss, L}$. 

Therefore, the mass loss rate at any time is the minimum of these two rates: $\dot{M}_\mathrm{atm} = \min[\dot{M}_\mathrm{L}, \dot{M}_\mathrm{B}]$. Typically, planets begin their evolution in the luminosity-limited regime due to their large $R_\mathrm{rcb}$, and transition into the Bondi-limited regime as the atmosphere contracts and the cooling of the core can become important. In the following section, we derive the cooling and mass loss time-scales analytically and show that cooling and contraction can preserve both the massive H/He-rich envelopes of sub-Neptunes as well as the light remnant primordial envelopes of super-Earths. 

\section{Analytic Results} \label{sec:analytic}
\subsection{Overview of results}
\begin{figure*}
\centering
\includegraphics[width=0.95\textwidth]{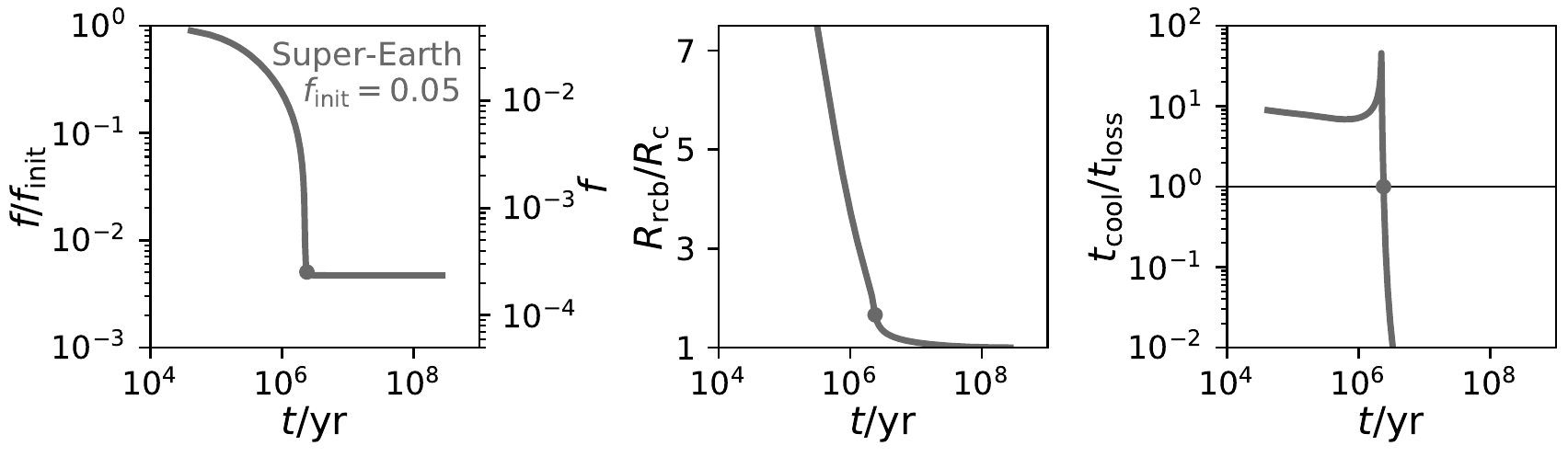}
\includegraphics[width=0.95\textwidth]{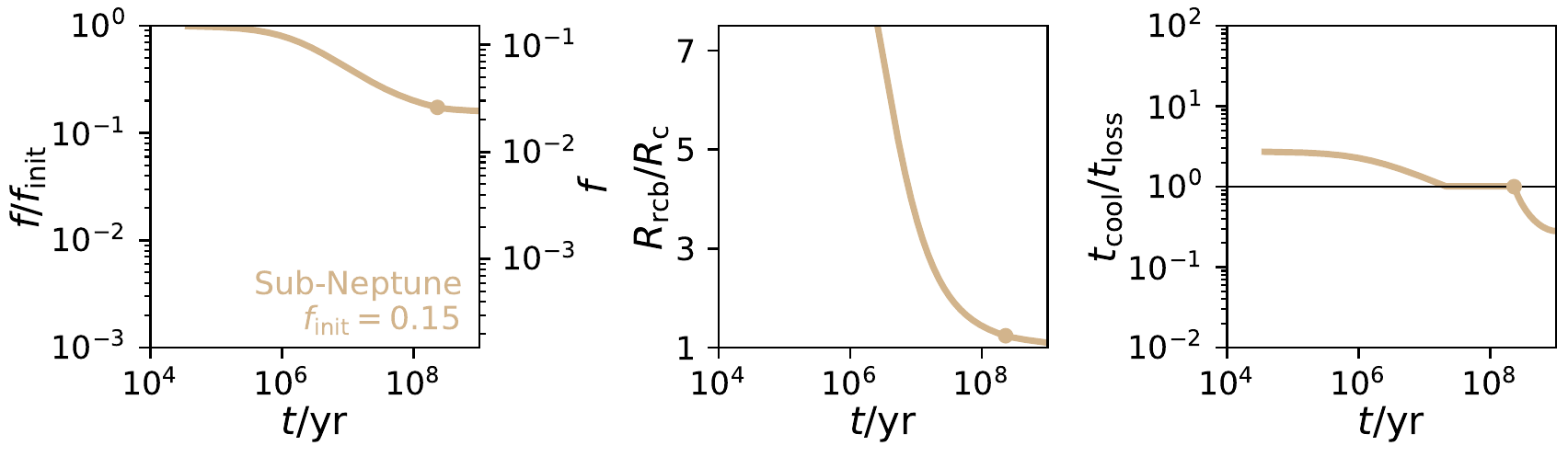}
\caption{Example of numerical simulations comparing the time evolution of atmospheric mass, $f$, radiative-convective boundary radius, $R_\mathrm{rcb}$, and ratio of the cooling time-scale to the mass loss time-scale, $t_\mathrm{cool}/t_\mathrm{loss}$, of two planets: one that evolves into a super-Earth (top panel) and one that remains a sub-Neptune (bottom panel). Both planets have the same mass, $M_\mathrm{c} = 4 M_\oplus$, and equilibrium temperature, $T_\mathrm{eq} = 1000\ \mathrm{K}$. The time at which $t_\mathrm{cool} = t_\mathrm{loss}$ in each simulation is marked by a dot. In the upper row, gray lines show the evolution of a planet with envelope mass at the time of disk dispersal $f_\mathrm{init} =0.05$. In the bottom row, tan lines show evolution of a planet with $f_\mathrm{init} = 0.15$. Both planets begin by undergoing spontaneous mass loss and shrinking until their radiative-convective boundaries are of order the core radius. At this transition point, the planet on the top row has sufficient thermal energy remaining in the core w.r.t. the envelope to power further mass loss. This mass loss proceeds until $f$ decreases enough to lower $t_\mathrm{cool}$ below $t_\mathrm{loss}$. This planet loses nearly all of its initial envelope and becomes a super-Earth, but it retains a thin primordial layer of hydrogen, $f_\mathrm{ret} \sim 10^{-4}$. Meanwhile, the planet on the bottom row ceases to lose mass by the end of the spontaneous mass loss phase, because the energy available for cooling is dominated by the envelop and not the core. The mass-loss is therefore terminated at the end of the spontaneous mass-loss phase and the cooling of the underlying core does not significantly effect the planet's evolution and the planet cools and contracts over gigayear time-scales. This specific planet retains 20\% of its initial envelope and corresponds to a sub-Neptune.} \label{fig:sim_comp_SESN_simple}
\end{figure*}

To give context to our analytic results and provide physical intuition for the processes that preserve residual atmospheres of super-Earths, we start by presenting the basic outcomes of two numerical simulations. These simulations are described in detail in Section \ref{sec:numerical}. Figure \ref{fig:sim_comp_SESN_simple} shows the evolution in time of the atmospheric mass, $f=M_\mathrm{atm}/M_\mathrm{c}$, the radiative-convective boundary, $R_\mathrm{rcb}$, and ratio of the cooling and mass loss time-scales, $t_\mathrm{cool}/t_\mathrm{loss}$, of two planets: one that will turn into a super-Earth and one that will remain a sub-Neptune. These two planets have the same physical parameters, except that the planet in the bottom row, i.e. the sub-Neptune, starts with a more massive primordial envelope. Immediately post-disk dispersal, which occurs instantaneously at $t=0$ in these simulations, the atmospheres quickly shed their loosely-bound outer layers, and both the atmospheric mass and radiative-convective boundary decrease. This is illustrated in the early times in both rows of Figure \ref{fig:sim_comp_SESN_simple}, where both planets lose $>50$\% of their initial envelopes in the first $10^6$ to $10^7$ years. Spontaneous mass loss continues until the mass loss becomes Bondi-limited rather than luminosity-limited, i.e., when the atmospheric cooling time-scale becomes less than the Bondi loss time-scale.

After this spontaneous mass loss phase the underlying rocky core, which is thermally coupled to the base of the atmosphere, may play a role in a planet's further evolution. As the core cools, it deposits thermal energy into the atmosphere. So long as the energy of the atmosphere itself dominates the available energy for cooling in the system, the core's cooling does not significantly affect the atmosphere's evolution. But if after spontaneous mass loss the core dominates the available cooling energy, then the cooling of the core can supply energy to the envelope as energy is lost by radiative diffusion across the radiative-convective boundary. This keeps the atmosphere inflated, which drives further mass loss. This effect was examined in detail in \citet{GSS16}, which found that the core's released thermal energy can be sufficient to unbind entire atmospheres. This is the core-powered mass loss mechanism, which can transform sub-Neptunes into super-Earths. The influence of the core is illustrated in Figure \ref{fig:sim_comp_SESN_simple}: after spontaneous mass loss, the atmosphere's available cooling energy is much larger than the core's thermal energy in the $f_\mathrm{init} = 0.15$ case (bottom). Therefore, the planet's cooling time-scale is nearly unaffected by the core and is already less than the mass loss time-scale from the contraction due to spontaneous mass loss alone. This planet will slowly cool and contract on gigayear time-scales without losing further atmospheric mass. These planets correspond to observed sub-Neptunes, and they typically retain $\sim 20$\% of their initial envelopes. Conversely, in the $f_\mathrm{init} = 0.05$ case (top), the core's thermal energy keeps the atmosphere inflated at $R_\mathrm{rcb} \sim 2 R_\mathrm{c}$, thereby lengthening the planet's cooling time-scale and enabling further mass loss. This planet loses most of its atmosphere and becomes a super-Earth \citep{GSS18,GS19}.

While super-Earths are produced under the core-powered mass loss mechanism, Figure \ref{fig:sim_comp_SESN_simple} shows that their primordial envelopes are not necessarily totally stripped. If the hot core can eventually cool on a shorter time-scale than mass loss occurs, the atmosphere will contract and cut off any further loss in the same fashion as the sub-Neptunes. Here, we briefly describe the physics of this process, which we derive more thoroughly in the remainder of the section.

The two basic physical quantities that evolve with time in the system are the atmospheric mass, $M_\mathrm{atm} \equiv f M_\mathrm{c}$, and the radius of the radiative-convective boundary, $R_\mathrm{rcb}$. After the atmosphere transitions into the Bondi loss regime, the mass loss rate is set by the density of the outflow: $\dot{M}_\mathrm{atm} \propto \rho_\mathrm{rcb}$ (see Equation \ref{eq:bondiloss}). This density depends linearly on the mass of the atmosphere: $\rho_\mathrm{rcb} \propto M_\mathrm{atm}$ (see Equation \ref{eq:mass_int}). Therefore, the mass loss time-scale, $t_\mathrm{loss} = M_\mathrm{atm}/\dot{M}_\mathrm{atm}$, is independent of the atmospheric mass. However, the planet's cooling time-scale, $t_\mathrm{cool} = E/\dot{E}$, is not independent of the atmospheric mass. The core's thermal energy is independent of the atmospheric mass. But as mass is lost at nearly constant radius, the density of the atmosphere, and therefore $\rho_\mathrm{rcb}$, decreases. Radiative diffusion across the radiative-convective boundary thus becomes more efficient, and the luminosity of the planet increases: $L = \dot{E} \propto 1/\rho_\mathrm{rcb} \propto 1/M_\mathrm{atm}$ (see Equation \ref{eq:luminosity}). In this way, as the atmospheric mass decreases, the cooling time-scale, $t_\mathrm{cool} \propto M_\mathrm{atm}$, also decreases. In summary, as $M_\mathrm{atm}$ decreases at nearly constant $R_\mathrm{rcb}$, $t_\mathrm{loss}$ is constant, while $t_\mathrm{cool}$ decreases. Thus there exists an atmospheric mass, which we term $f_\mathrm{ret}$, at which the cooling time-scale will become shorter than the mass loss time-scale. Once $f$ decreases to this value, the envelope will quickly contract and cut off further mass loss, preserving the remaining primordial gas. This cross-over occurs for the $f_\mathrm{init} = 0.05$ case that we show on the top row of Figure \ref{fig:sim_comp_SESN_simple} at the time marked with dots on each panel. In this case, the core cooling time-scale $t_\mathrm{cool}$ becomes shorter than $t_\mathrm{loss}$ when $f/f_\mathrm{init} \sim 5 
\times 10^{-3}$, after which no further mass is lost. This super-Earth then cools and contracts at constant atmospheric mass, just as the sub-Neptune does.

Our aim in this section is to quantify the residual H/He gas we expect these planets to retain after core-powered mass loss, $f_\mathrm{ret}$. To find this mass fraction, we analytically solve for the atmospheric mass at which the planet's cooling rate becomes faster than its mass loss rate. In the remainder of this section, we divide the atmospheric evolution into two tractable analytic regimes. First we quantify the spontaneous mass loss mechanism in Section \ref{sec:sml}. In this mass loss phase, we focus on the envelope's own energy as the main source of available cooling energy. This regime sets the radiative-convective boundary radius and atmospheric mass for the next phase of evolution. After spontaneous mass loss phase, the remaining gravitational and thermal energy of the atmosphere can exceed the thermal energy of the core. In this case, the atmosphere contracts and cools without further significant mass loss, and the planet remains a sub-Neptune. On the other hand, if the underlying core's thermal energy is larger than the atmospheric energy, the core's cooling will play a dominant role in the evolution of the atmosphere and the resulting atmospheric loss can turn a sub-Neptune into a super-Earth. In this case, presented in Section \ref{sec:coreE}, we focus on the core's thermal energy as main cooling energy source.

\subsection{Spontaneous mass loss}\label{sec:sml}
When the disk disperses, the planet will cool by radiative diffusion across the $R_\mathrm{rcb}$ with a cooling time-scale, $t_\mathrm{cool} \equiv |E/L|$. Here $L$ is the luminosity of the envelope expressed in Equation \ref{eq:luminosity}. While the atmosphere extends significantly, $R_\mathrm{rcb} \gg R_\mathrm{c}$, the adiabatic gradient of the atmosphere keeps the core nearly constant in temperature as the atmosphere contracts. Thus, we neglect any energy contribution from the core's cooling in this stage. The energy available for cooling the envelope is therefore its integrated thermal and gravitational potential energy as expressed in Equation \ref{eq:energy_int}. Assuming that the energy is concentrated in the interior of the convective region (true for $\gamma < 3/2$), this integral can be approximated as \citep[see][Eq 10]{GSS16}
\begin{equation}\label{eq:energygeneral}
    E_\mathrm{atm} \simeq -\frac{(\gamma-1)^2}{\gamma (3-2\gamma)} G M_\mathrm{c} 4\pi R_\mathrm{c}^2 \rho_\mathrm{rcb} \bigg(\frac{R_\mathrm{B}'}{R_\mathrm{c}}\bigg)^{\frac{1}{\gamma-1}} \mathrm{.}
\end{equation}
The density at the $R_\mathrm{rcb}$, $\rho_\mathrm{rcb}$, can be expressed in terms of the atmospheric mass, $M_\mathrm{atm}$, by integrating the density profile given in Equation \ref{eq:density_struc}. We approximate this integral in this `thick' regime as \citep[see][Eq 11]{GSS16}
\begin{equation}\label{eq:atmmass}
    M_\mathrm{atm} \simeq A 4 \pi R_\mathrm{rcb}^3 \rho_\mathrm{rcb} \bigg(\frac{R_\mathrm{B}'}{R_\mathrm{rcb}}\bigg)^{\frac{1}{\gamma-1}} \mathrm{,}
\end{equation}
where $A$ is an integration constant equal to $5\pi/16$ in the limit $R_\mathrm{rcb} \gg R_\mathrm{c}$.
This expression is derived under the assumption that $R_\mathrm{c} \ll R_\mathrm{rcb} \lesssim R_\mathrm{B}'$, $\gamma > 4/3$, and the mass in the isothermal region is negligible. The first approximation will begin to fail as $R_\mathrm{rcb}$ approaches $R_\mathrm{c}$, which will be treated in the thin regime section below.
Thus the density at $R_\mathrm{rcb}$ is
\begin{equation}\label{eq:rhorcb_est}
    \rho_\mathrm{rcb} \simeq \frac{M_\mathrm{atm}}{A 4\pi R_\mathrm{B}'^{\frac{1}{\gamma-1}} R_\mathrm{rcb}^{3-\frac{1}{\gamma-1}}} \mathrm{.}
\end{equation}
This approximation allows us to simplify Equation \ref{eq:energygeneral} as 
\begin{equation}\label{eq:energysimple}
    E_\mathrm{atm} \simeq -\frac{(\gamma - 1)^2 G M_\mathrm{c} M_\mathrm{atm}}{A \gamma (3-2\gamma) R_\mathrm{c}} \bigg(\frac{R_\mathrm{c}}{R_\mathrm{rcb}}\bigg)^{3-\frac{1}{\gamma-1}} \mathrm{.}
\end{equation}
Combining this expression with Equation \ref{eq:luminosity}, we derive the cooling time-scale in the regime for which the atmospheric potential energy dominates the energy available in the system:
\begin{equation}\label{eq:tcool}
\begin{split}
t_\mathrm{cool} &\simeq  K_\mathrm{cool}(\gamma) f^2 M_\mathrm{c}^{2- \frac{1}{\gamma-1}} R_\mathrm{c}^{2-\frac{1}{\gamma-1}} T_\mathrm{eq}^{-3+\frac{1}{\gamma-1}} R_\mathrm{rcb}^{-2(3-\frac{1}{\gamma-1})} \\
&\simeq 6.80 \times 10^{14} \bigg(\frac{f}{0.05}\bigg)^2 \bigg(\frac{M_\mathrm{c}}{3 M_\oplus}\bigg)^{-7/8} \bigg(\frac{T_\mathrm{eq}}{1000\ \mathrm{K}}\bigg)^{-1/2}  \bigg(\frac{R_\mathrm{rcb}}{R_\mathrm{c}}\bigg)^{-1} \mathrm{\ s}
\end{split} \mathrm{,}
\end{equation}
where $K_\mathrm{cool}(\gamma) = 3 \kappa (\gamma - 1)^2 G/(256\pi^2 A^2 \gamma (3-2\gamma)\sigma) [(\gamma-1) G \mu/(\gamma k_\mathrm{B})] ^{-1-\frac{1}{\gamma-1}}$, and we have evaluated the second line using $\gamma=7/5$, an adiabatic index appropriate for an atmosphere of diatomic hydrogen. 

To assess whether the mass loss rate is limited initially by the internal luminosity or the hydrodynamic outflow, we compare the luminosity-limited mass loss time-scale to the Bondi mass loss time-scale. The luminosity-limited mass loss time-scale, $t_\mathrm{loss, L} \equiv |M_\mathrm{atm}/\dot{M}_\mathrm{L}| \simeq E_\mathrm{atm}/L$, is, to first order, equivalent to the cooling time-scale. The Bondi mass loss time-scale, $t_\mathrm{loss, B} \equiv |M_\mathrm{atm}/\dot{M}_\mathrm{B}|$, is, combining Equations \ref{eq:bondiloss} and \ref{eq:atmmass}

\begin{equation}\label{eq:tloss}
\begin{split}
    t_\mathrm{loss, B} &\simeq  K_\mathrm{loss}(\gamma) M_\mathrm{c}^{- 2 + \frac{1}{\gamma-1}} T_\mathrm{eq}^{\frac{3}{2}-\frac{1}{\gamma-1}} R_\mathrm{rcb}^{3-\frac{1}{\gamma-1}} \exp{[R_\mathrm{B}/R_\mathrm{rcb}]}\\
    &\simeq 421 \bigg(\frac{M_\mathrm{c}}{3 M_\oplus}\bigg)^{5/8} \bigg(\frac{T_\mathrm{eq}}{1000\ \mathrm{K}}\bigg)^{-1} \bigg(\frac{R_\mathrm{rcb}}{R_\mathrm{c}}\bigg)^{1/2} \exp{[R_\mathrm{B}/R_\mathrm{rcb}]} \mathrm{\ s} \mathrm{,}
\end{split}
\end{equation}
where $K_\mathrm{loss}(\gamma) = A/\mathrm{e} (k_\mathrm{B}/ \mu)^{\frac{3}{2} - \frac{1}{\gamma-1}} G^{- 2 + \frac{1}{\gamma-1}}$ and the second line is again evaluated for $\gamma=7/5$. The loss time-scale is much shorter than the cooling time-scale initially, when $R_\mathrm{rcb} \gg R_\mathrm{c}$. Therefore, atmospheres lose mass at the luminosity-limited mass loss rate: $\dot{M}_\mathrm{atm} = \dot{M}_\mathrm{L}$ and $t_\mathrm{loss} = t_\mathrm{cool}$. However, as the planet cools and loses mass, the atmosphere contracts. The Bondi mass loss time-scale is independent of $f$, and as $R_\mathrm{rcb}$ decreases, this time-scale increases exponentially and can become longer than the luminosity-limited time-scale, $t_\mathrm{cool} \propto R_\mathrm{rcb}^{-1} f^2$. Once this occurs, the atmospheric mass loss rate transitions from $\dot{M}_\mathrm{L}$ to $\dot{M}_\mathrm{B}$. This change in mass loss rate has important consequences for the future of the planet: rather than the loss rate increasing as mass is lost, which would cause runaway atmospheric loss, the mass loss rate is independent of the atmospheric mass and exponentially decreases as the planet contracts, which can save the remaining atmosphere.

We quantify the crossover point when the Bondi loss rate decreases below the luminosity-limited rate by setting the two time-scales equal. Since the loss time-scale is independent of $f$ and the cooling time-scale depends only weakly on it compared to the exponential dependence on $R_\mathrm{rcb}$ in the loss time-scale, we take the mass as roughly constant for this analytic derivation and set $t_\mathrm{loss, B} = t_\mathrm{cool}$ to solve for $R_\mathrm{rcb}$. This value is the radiative-convective boundary to which the atmosphere must shrink for the Bondi mass loss rate to become equal to the rate limiting atmospheric loss.

Equating Equations \ref{eq:tcool} and \ref{eq:tloss} yields
\begin{equation}\label{eq:exprcb}
    R_\mathrm{rcb}^{n} \simeq Z \exp{[-R_\mathrm{B}/R_\mathrm{rcb}]},
\end{equation}
where $n \equiv 3(3-1/(\gamma-1))$ and $Z = K_\mathrm{cool} /K_\mathrm{loss} f^2 M_\mathrm{c}^{4-\frac{2}{\gamma-1}} R_\mathrm{c}^{2-\frac{1}{\gamma-1}} T_\mathrm{eq}^{-\frac{9}{2}+\frac{2}{\gamma-1}}$. This equation has no exact solution for $R_\mathrm{rcb}$ in terms of elementary functions. However, in the limit where $R_\mathrm{B}/R_\mathrm{rcb}$ is large, the exponential term on the right-hand side of Equation \ref{eq:exprcb} changes much faster than the power-law term on the left-hand side with changing $R_\mathrm{rcb}$. We therefore approximate the power-law term as a constant: $R_\mathrm{rcb}^n \approx (\epsilon R_\mathrm{B})^n$. This approach requires an estimate of the expected final result, so we take $\epsilon \approx 0.03$. Varying the value of $\epsilon$ by a factor of a few does not have a large effect on the solution, however.

Solving for the remaining $R_\mathrm{rcb}$ term in the exponential term of Equation \ref{eq:exprcb} yields the critical radius $R_\mathrm{rcb}$ at which $t_\mathrm{cool} = t_\mathrm{loss, B}$ 
\begin{equation}\label{eq:rcbapprox}
\begin{split}
    R_\mathrm{rcb} &\simeq \frac{-R_\mathrm{B}}{\ln{[\epsilon^n R_\mathrm{B}^n/Z]}} \\
    \frac{R_\mathrm{rcb}}{R_\mathrm{c}} &\approx \cfrac{38.0 \times \bigg(\cfrac{M_\mathrm{c}} {3 M_\oplus}\bigg)^{3/4} \bigg(\cfrac{T_\mathrm{eq}} {1000 \mathrm{\ K}}\bigg)^{-1}}{27.9 -1.5\ln{\bigg[\cfrac{\epsilon}{0.03}\bigg]} + 2\ln{\bigg[\cfrac{f} {0.05}\bigg]} +2\ln{\bigg[\cfrac{T_\mathrm{eq}} {1000 \mathrm{\ K}}\bigg]} - 2.625\ln{\bigg[\cfrac{M_\mathrm{c}} {3 M_\oplus}\bigg]}}
\end{split} \mathrm{,}
\end{equation}
where we have again taken $\gamma=7/5$ in the last line.

Using this new $R_\mathrm{rcb}$, we estimate the mass lost in this initial spontaneous mass loss phase. The mass that was initially contained between this new radius and the initial $R_\mathrm{rcb}$ is taken to be lost in this phase. An analytic prediction can thus be made for $M_\mathrm{lost}$ by approximating the density profile in Equation \ref{eq:density_struc} as $\rho_\mathrm{init}(r) \approx \rho_\mathrm{rcb} (R_\mathrm{B}'/r)^{1/(\gamma-1)}$.
This simplification of the $r$ dependence allows the mass lost to be integrated analytically:
\begin{equation}\label{eq:mlostanalytic}
    \frac{f_\mathrm{lost}}{f_\mathrm{init}} \simeq \frac{R_\mathrm{rcb,init}^{3-1/(\gamma-1)} - R_\mathrm{rcb,crit}^{3-1/(\gamma-1)}}{ R_\mathrm{rcb,init}^{3-1/(\gamma-1)} - R_\mathrm{c}^{3-1/(\gamma-1)}} \mathrm{,}
\end{equation}
where $R_\mathrm{rcb,init} = R_\mathrm{B}'$.
The mass lost found is typically near 75\% of the planet's initial captured atmosphere, which is consistent results given in \citet{GSS16} for $\gamma = 7/5$.

\subsection{Core cooling regime}\label{sec:coreE}
At the end of the spontaneous mass loss phase, mass loss stops being limited by the planet's cooling luminosity and becomes instead limited by the rate at which gas molecules can escape at the Bondi radius. Additionally, since the radius of the envelope is of order the core's radius, the thermal energy available for cooling in the hot underlying core, $E_\mathrm{c}$, can be released and can contribute to the further evolution of the atmosphere. If $E_\mathrm{c} \lesssim |E_\mathrm{atm}|$, then the thermal energy of the core is insufficient to significantly affect the atmosphere's evolution. Now, any slight atmospheric contraction causes the mass loss time-scale to increase exponentially. Meanwhile, the cooling time-scale, previously coupled to the mass loss time-scale, does not substantially change even with the core's additional contribution. Thus the mass loss time-scale quickly becomes longer than the planet's cooling time-scale. Such planets will slowly cool and contract on gigayear time-scales and will lose no more atmospheric mass. These planets correspond to observed sub-Neptunes.

Conversely, if $E_\mathrm{c} \gtrsim |E_\mathrm{atm}|$, then there is more energy available for cooling than we considered above. Therefore, $t_\mathrm{cool} \propto E$ is still longer than $t_\mathrm{loss}$, so mass loss continues until the core has cooled sufficiently for the envelope to contract. In what follows in this section, we assume that the core's thermal energy dominates the total energy available for cooling. We use this to determine a planet's further evolution by again setting the cooling and mass loss time-scales equal, as above.

The core's thermal energy available for cooling is determined by the temperature of the base of the atmosphere, because the core and atmosphere are thermally coupled (see Equation \ref{eq:energycore}). The temperature at the base of the atmosphere is in turn is related to the temperature at the radiative-convective boundary via the adiabatic profile of the envelope. If $R_\mathrm{rcb} \gtrsim 2 R_\mathrm{c}$, the temperature at the base of the atmosphere, and thus the core temperature, $T_\mathrm{c} = T(R_\mathrm{c})$, varies little as $R_\mathrm{rcb}$ changes. Therefore we approximate the core temperature, in this regime, as independent to first order of $R_\mathrm{rcb}$: $T_\mathrm{c} \simeq T_\mathrm{eq} R_\mathrm{B}'/R_\mathrm{c}$. Once $R_\mathrm{rcb} \lesssim 2 R_\mathrm{c}$, however, the temperature at the core-atmosphere boundary does decrease substantially, thereby releasing thermal energy into the envelope, as the radiative-convective boundary approaches the core \citep[see][Eq 19]{GSS16}:
\begin{equation}
    T_\mathrm{c}(R_\mathrm{rcb} < 2 R_\mathrm{c}) \simeq T_\mathrm{eq} \frac{R_\mathrm{B}' \Delta R_\mathrm{a}}{R_\mathrm{c}^2} \mathrm{,}
\end{equation}
where we define the width of the atmosphere $\Delta R_\mathrm{a} \equiv R_\mathrm{rcb} - R_\mathrm{c}$. We note here that the core temperature does not depend on the mass of the atmosphere in either regime. 

Inserting these temperatures into Equation \ref{eq:energycore}, we derive the core energy available for cooling relevant for the end of a planet's evolution when the cooling time-scales and mass-loss time-scales can become comparable:
\begin{equation}\label{eq:Ecorethin}
    E_\mathrm{c} \simeq \begin{cases} \cfrac{\gamma - 1}{\gamma (\gamma_\mathrm{c} -1)} \cfrac{\mu}{\mu_\mathrm{c}} \cfrac{G M_\mathrm{c}^2}{R_\mathrm{c}} &\mbox{if } R_\mathrm{rcb} > 2 R_\mathrm{c} \\
    \cfrac{\gamma-1}{\gamma (\gamma_\mathrm{c} -1)} \cfrac{\mu}{\mu_\mathrm{c}} \cfrac{G M_\mathrm{c}^2}{R_\mathrm{c}} \cfrac{\Delta R_\mathrm{a}} {R_\mathrm{c}} & \mbox{if } R_\mathrm{rcb} \lesssim 2 R_\mathrm{c} \end{cases}  \mathrm{.}
\end{equation}
As the core cools in tandem with the base of the atmosphere, it releases its thermal energy into the envelope. This energy input impedes further atmospheric contraction, thereby sustaining mass loss until the planet's core has cooled significantly.

If $R_\mathrm{rcb} \gtrsim 2 R_\mathrm{c}$, we still approximate the atmospheric mass by Equation \ref{eq:atmmass}. However, the integration factor, $A$, is no longer necessarily approximately equal to $5\pi/16$, so we now use its integral form
\begin{equation}
    A \simeq \int_{R_\mathrm{c}/R_\mathrm{rcb}}^1 x^2 \bigg(\frac{1}{x}-1\bigg)^{\frac{1}{\gamma-1}} \mathrm{d}x \mathrm{,}
\end{equation}
where $x \equiv r/R_\mathrm{rcb}$. This integration factor is smaller than $5\pi/16$ and decreases as the atmosphere contracts.

If the atmosphere is sufficiently thin, $R_\mathrm{c} \lesssim 2 R_\mathrm{rcb}$, the atmospheric mass is instead better approximated as
\begin{equation}\label{eq:Matmthin}
    M_\mathrm{atm} \simeq \frac{\gamma-1}{\gamma} 4 \pi R_\mathrm{c}^2 \Delta R_\mathrm{a} \rho_\mathrm{rcb} \bigg(\frac{R_\mathrm{B}' \Delta R_\mathrm{a}}{R_\mathrm{c}^2} \bigg)^{\frac{1}{\gamma-1}} \mathrm{.}
\end{equation}

Since the core is thermally coupled to the atmosphere, its cooling rate is limited by the rate of radiative diffusion across the radiative-convective boundary of the envelope (see Equation \ref{eq:luminosity}). In the core cooling regime, the cooling time-scale can therefore be approximated as the ratio of the core's thermal energy and the luminosity, $t_\mathrm{cool} = |E_\mathrm{c}/L|$. For $R_\mathrm{rcb} \gtrsim 2 R_\mathrm{c}$, the cooling time-scale is
\begin{equation}
\begin{split}
    t_\mathrm{cool} &\simeq \frac{C_\mathrm{cool}(\gamma)}{A} f M_\mathrm{c}^{2 -\frac{1}{\gamma-1}} R_\mathrm{c}^{-1} T_\mathrm{eq}^{-3+\frac{1}{\gamma-1}} R_\mathrm{rcb}^{-3 + \frac{1}{\gamma-1}}\\
    &\simeq 7.2 \times 10^{14} \bigg(\frac{f} {0.01}\bigg)
    \bigg(\frac{M_\mathrm{c}}{3 M_\oplus}\bigg)^{-7/8} \bigg(\frac{T_\mathrm{eq}}{1000\ \mathrm{K}}\bigg)^{-1/2}  \bigg(\frac{R_\mathrm{rcb}}{R_\mathrm{c}}\bigg)^{-1/2} \mathrm{\ s,}
\end{split}
\end{equation}
while for $R_\mathrm{rcb} \lesssim 2 R_\mathrm{c}$, the dependence on the radiative-convective boundary changes to
\begin{equation}\label{eq:tcool_thin}
\begin{split}
    t_\mathrm{cool} &\simeq C_\mathrm{cool}(\gamma) \frac{\gamma}{\gamma-1} f M_\mathrm{c}^{2-\frac{1}{\gamma-1}} R_\mathrm{c}^{-4+\frac{2}{\gamma-1}} T_\mathrm{eq}^{-3+\frac{1}{\gamma-1}} \Delta R_\mathrm{a}^{-\frac{1}{\gamma-1}}\\
    &\simeq 5.05 \times 10^{14} \bigg(\frac{f}{0.01}\bigg) \bigg(\frac{M_\mathrm{c}}{3 M_\oplus}\bigg)^{-7/8} \bigg(\frac{T_\mathrm{eq}}{1000 \mathrm{\ K}}\bigg)^{-1/2} \bigg(\frac{\Delta R_\mathrm{a}}{R_\mathrm{c}}\bigg)^{-5/2} \mathrm{\ s,} 
\end{split}
\end{equation}
where $C_\mathrm{cool}(\gamma) = 3 \kappa k_\mathrm{B}/(256 \pi^2 \mu_\mathrm{c} \sigma (\gamma_\mathrm{c}-1))[(\gamma-1) G \mu /(\gamma k_\mathrm{B})]^{-1/(\gamma-1)}$. In both cases we evaluate the second line for $\gamma=7/5$.

The mass loss time-scale, now in the Bondi-limited regime after spontaneous mass loss, remains as given in Equation \ref{eq:tloss} for $R_\mathrm{rcb} \gtrsim 2 R_\mathrm{c}$. For $R_\mathrm{rcb} \lesssim 2 R_\mathrm{c}$, we can solve for the mass loss time-scale using Equations \ref{eq:bondiloss} and \ref{eq:Matmthin}:
\begin{equation}\label{eq:tloss_thin}
\begin{split}
    t_\mathrm{loss} &\simeq C_\mathrm{loss}(\gamma) M_\mathrm{c}^{-2+\frac{1}{\gamma-1}} R_\mathrm{c}^{2-\frac{2}{\gamma-1}} T_\mathrm{eq}^{\frac{3}{2}- \frac{1}{\gamma-1}} \Delta R_\mathrm{a}^{1+\frac{1}{\gamma-1}} \exp{[R_\mathrm{B}/(R_\mathrm{c}+\Delta R_\mathrm{a})]} \\ 
    &\simeq 122 \bigg(\frac{T_\mathrm{eq}}{1000 \mathrm{\ K}}\bigg)^{-1} \bigg(\frac{M_\mathrm{c}}{3 M_\oplus}\bigg)^{5/8} \bigg(\frac{\Delta R_\mathrm{a}}{R_\mathrm{c}}\bigg)^{7/2} \exp{[R_\mathrm{B}/(R_\mathrm{c}+\Delta R_\mathrm{a})]} \mathrm{\ s,}
\end{split}
\end{equation}
where $C_\mathrm{loss}(\gamma) = [(\gamma-1)/\gamma]^{1+1/(\gamma-1)} (k_\mathrm{B}/\mu)^{3/2-1/(\gamma-1)} G^{-2 + 1/(\gamma-1)}$
. For typical parameters, loss time-scales can be short during this late stages of evolution ($\approx 700$ years for $M_\mathrm{c} =3 M_\oplus$, $T_\mathrm{eq}=1000$ K, and $R_\mathrm{rcb} = 2 R_\mathrm{c}$), but the time-scales depend exponentially on the core mass and the inverse of the equilibrium temperature. This analytically-derived mass loss time-scale is independent of the atmospheric mass $f$, as $\dot M_\mathrm{atm}$ depends linearly on the atmospheric mass and hence cancels out.

\begin{figure*}
\centering
    \includegraphics[width=0.99\textwidth]{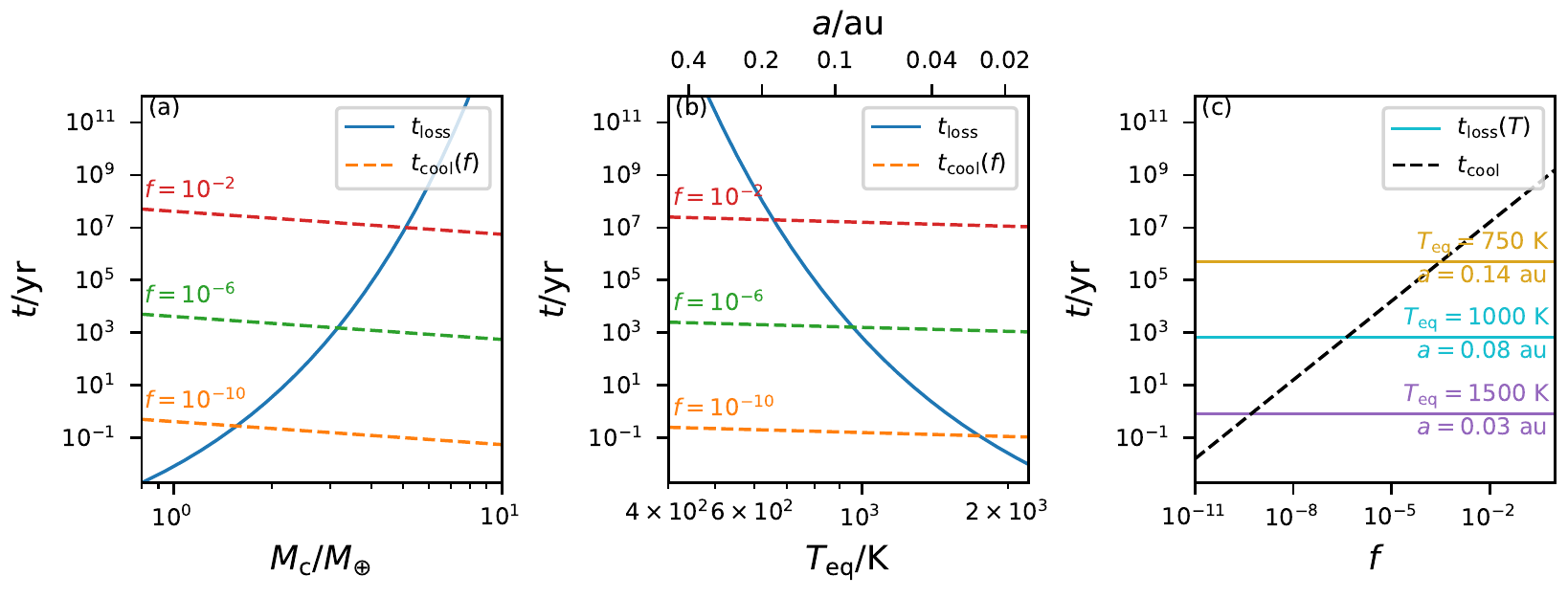}
    \caption{Comparison of the analytically derived thin regime mass loss and cooling time-scales, $t_\mathrm{loss}$ and $t_\mathrm{cool}$ respectively, for planets towards the end of their evolution with $R_\mathrm{rcb}=2 R_\mathrm{c}$. Panel (a) shows the time-scales as functions of core mass, $M_\mathrm{c}$, at fixed equilibrium temperature, $T_\mathrm{eq}=1000$ K. Panel (b) displays how these time-scales depend on $T_\mathrm{eq}$ at fixed $M_\mathrm{c} = 3 M_\oplus$. We also show a conversion of equilibrium temperature to distance from a star of solar luminosity, $a$, via Equation \ref{eq:tempa}. Finally, in panel (c) we plot $t_\mathrm{cool}$ and $t_\mathrm{loss}$ as a function of atmospheric mass fraction, $f$, at fixed $M_\mathrm{c} = 3 M_\oplus$. All time-scales are evaluated using $\gamma=7/5$. Only one line is plotted for $t_\mathrm{loss}$ in (a) and (b), as $t_\mathrm{loss}$ is independent of $f$ (see Equation \ref{eq:tloss_thin}). Similarly, $t_\mathrm{cool}$ is nearly independent of $T_\mathrm{eq}$ (see Equation \ref{eq:tcool_thin}, $t_\mathrm{cool} \propto T_\mathrm{eq}^{-1/2}$), such that the lines are indistinguishable on the scale shown in (c). The intersections between the lines yield the set of parameters for which $t_\mathrm{cool}=t_\mathrm{loss}$. As shown in (a), for a planet at given $T_\mathrm{eq}$, $t_\mathrm{loss}$ increases exponentially with increasing $M_\mathrm{c}$, while $t_\mathrm{cool}$ is nearly independent of $M_\mathrm{c}$ but is proportional to $f$. Thus as $M_\mathrm{c}$ is increased, the $f$ value for which the two lines intersect increases exponentially. Similarly, (b) shows that this critical $f$ decreases exponentially with increasing $T_\mathrm{eq}$. In (c), we demonstrate the final atmospheric mass fraction, $f_\mathrm{ret}$, at which $t_\mathrm{cool}= t_\mathrm{loss}$ increases for decreasing $T_\mathrm{eq}$, which corresponds to increasing $t_\mathrm{loss}$ values since the loss time-scale for a given $T_\mathrm{eq}$ is independent of $f$. Taken together, these results show that as a planet of given core mass and equilibrium temperature loses atmosphere, its loss time-scale is nearly constant, and its cooling time-scale decreases. Thus the planet will encounter a critical $f$ at which $t_\mathrm{cool}= t_\mathrm{loss}$. This $f_\mathrm{ret}$ is exponentially dependent on $M_\mathrm{c}$ and $T_\mathrm{eq}$ and so varies widely over typical super-Earth values, and once a planet reaches this envelope fraction, mass loss will quickly halt as the planet cools and contracts.
}
\label{fig:time-scale_comps}
\end{figure*}

The results for the mass loss and cooling time-scales are compared in Figure \ref{fig:time-scale_comps}, varying $M_\mathrm{c}$, $T_\mathrm{eq}$, and $f$ in (a), (b), and (c) respectively, with fixed $R_\mathrm{rcb} = 2 R_\mathrm{c}$. This figure shows that $t_\mathrm{cool}$ decreases as $f$ decreases and is only weakly dependent on other planetary parameters. Conversely, $t_\mathrm{loss}$ is independent of $f$ but depends exponentially on $M_\mathrm{c}$ and $T_\mathrm{eq}$ as shown in Equation \ref{eq:tloss_thin}, so this time-scale varies widely over typical super-Earth parameters. As shown in Figure \ref{fig:time-scale_comps}, as a planet of given $M_\mathrm{c}$ and $T_\mathrm{eq}$ evolves and loses mass, it will naturally reach a critical atmospheric mass $f_\mathrm{ret}$ for which $t_\mathrm{cool}= t_\mathrm{loss}$, shown by the solid and dashed lines intersecting. That these two time-scales intersect is guaranteed because a planet starts out with $t_\mathrm{loss}< t_\mathrm{cool}$, but as it loses atmospheric mass $t_\mathrm{cool}$ deceases while the mass-loss rate remains almost unchanged (see panel (c) in Figure \ref{fig:time-scale_comps}). This allows $t_\mathrm{cool}$ to catch up with $t_\mathrm{loss}$ at a critical atmospheric mass, $f_\mathrm{ret}$. Once a planet reaches this critical atmospheric mass cooling becomes faster than loss. The core will cool, allowing the atmosphere to contract further. This contraction exponentially lowers the mass loss rate, shutting off mass loss. In this way, we expect this mass $f_\mathrm{ret}$ to be the final primordial envelope mass these planets retain, absent other factors affecting mass loss. The value of $f_\mathrm{ret}$ which balances the time-scales varies by orders of magnitude from $10^{-2}$ to $10^{-10}$ over typical super-Earth parameters.

\begin{figure*}
\centering
    \includegraphics[width=0.99\textwidth]{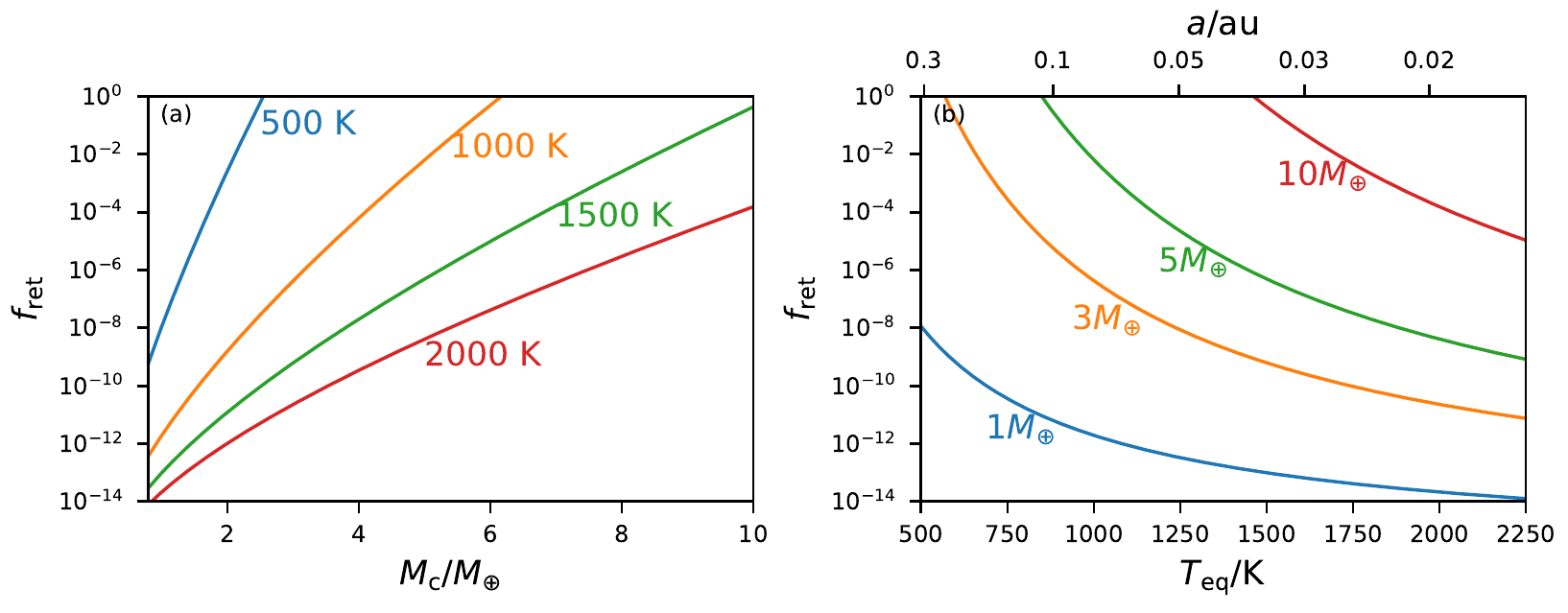}
\caption{Analytical estimates of the atmospheric mass $f_\mathrm{ret}$ retained after core-powered mass loss in the thin regime as a function of (a) core mass, $M_\mathrm{c}$, and (b) equilibrium temperature, $T_\mathrm{eq}$. In panel (a) we plot lines corresponding to four equilibrium temperatures spanning typical super-Earth environments, whereas in (b) we show four typical core masses. These values are calculated using Equation \ref{eq:f_thin}, with $\gamma=7/5$ and $R_\mathrm{rcb} = 2 R_\mathrm{c}$. $f_\mathrm{ret}$ varies exponentially over the super-Earth regime, with negligibly small envelopes retained for the least massive and hottest planets and thicker envelopes retained for cooler and more massive planets.}
\label{fig:f_thin}
\end{figure*}

We can analytically solve for this intersection point by setting the the mass loss and cooling time-scales equal. This is the atmospheric mass we expect these planets to retain after core-powered mass loss. Solving for $R_\mathrm{rcb} \gtrsim 2 R_\mathrm{c}$:
\begin{equation}\label{eq:f_core_thick}
\begin{split}
     f_\mathrm{ret} &\simeq \frac{K_\mathrm{loss} A}{C_\mathrm{cool}} M_\mathrm{c}^{-4+ \frac{2}{\gamma-1}} R_\mathrm{c} T_\mathrm{eq}^{\frac{9}{2} -\frac{2}{\gamma-1}} R_\mathrm{rcb}^{2(3 - \frac{1}{\gamma-1})} \exp{[R_\mathrm{B}/R_\mathrm{rcb}]}\\
    \frac{f_\mathrm{ret}}{0.01} &\simeq 1.2 \times 10^{-13}
    \bigg(\frac{M_\mathrm{c}}{3 M_\oplus}\bigg)^{3/2} 
    \bigg(\frac{T_\mathrm{eq}}{1000 \mathrm{\ K}}\bigg)^{-1/2} \bigg(\frac{R_\mathrm{rcb}} {R_\mathrm{c}}\bigg) \exp{[R_\mathrm{B}/R_\mathrm{rcb}]}
\end{split}
\end{equation}
and for $R_\mathrm{rcb} \lesssim 2 R_\mathrm{c}$: 
\begin{equation}\label{eq:f_thin}
\begin{split}
    f_\mathrm{ret} &\simeq \frac{C_\mathrm{loss}}{C_\mathrm{cool}} \frac{\gamma-1}{\gamma} M_\mathrm{c}^{-4+\frac{2}{\gamma-1}} R_\mathrm{c}^{6-\frac{4}{\gamma-1}} T_\mathrm{eq}^{\frac{9}{2}- \frac{2}{\gamma-1}} \Delta R_\mathrm{a}^{1+\frac{2}{\gamma-1}} \exp{[R_\mathrm{B}/(R_\mathrm{c}+\Delta R_\mathrm{a})]} \\
    \frac{f_\mathrm{ret}}{0.01} &\simeq 2.43 \times 10^{-13} \bigg(\frac{M_\mathrm{c}}{3 M_\oplus}\bigg)^{3/2} \bigg(\frac{T_\mathrm{eq}}{1000 \mathrm{\ K}}\bigg)^{-1/2} \bigg(\frac{\Delta R_\mathrm{a}}{R_\mathrm{c}}\bigg)^{6} \\ &\qquad
    \exp{\bigg[19.0  \bigg(\frac{M_\mathrm{c}}{3 M_\oplus}\bigg)^{3/4} \bigg(\frac{T_\mathrm{eq}}{1000 \mathrm{\ K}}\bigg)^{-1}} \bigg(\frac{R_\mathrm{c} + \Delta R_\mathrm{a}}{2 R_\mathrm{c}}\bigg)^{-1}\bigg]\mathrm{.}
\end{split}
\end{equation}
The solutions to Equation \ref{eq:f_thin} are shown in Figure \ref{fig:f_thin}, taking $R_\mathrm{rcb} = 2 R_\mathrm{c}$. As indicated by Figure \ref{fig:time-scale_comps}, $f_\mathrm{ret}$ is smaller for less massive, hotter planets, and larger for more massive, cooler planets. The mass fraction of this leftover H/He envelope ranges from negligible ($f \sim 10^{-10}$), to comparable to Earth's modern secondary atmosphere ($f \sim 10^{-6}$), all the way to many times more massive than the modern secondary atmosphere of Venus ($f > 10^{-4}$).

\subsection{Final pressures}
Given these final atmospheric masses, we can calculate the corresponding surface pressures. These pressures have implications for the chemistry possible near the surface, including the outgassing and final atmospheric compositions. The final surface pressure, once the atmosphere has contracted such that $R_\mathrm{rcb} \sim R_\mathrm{c}$, is
\begin{equation}\label{eq:pressure_thin}
    P_\mathrm{surf} \simeq \frac{G f M_\mathrm{c}^2}{4 \pi R_\mathrm{c}^{4}} \simeq 1 \mathrm{\ bar} \frac{f}{10^{-6}} \frac{M_\mathrm{c}}{M_\oplus} \mathrm{.}
\end{equation}
This approximation holds as the atmosphere contracts from $R_\mathrm{rcb} \sim$ a few $R_\mathrm{c}$ to a fully isothermal profile up to an order unity constant, which accounts for the appropriate atmospheric mass distribution.

\subsection{Optically thin limit}\label{sec:optthin}
Our assumptions for the luminosity of the atmosphere only hold if the atmosphere is optically thick, i.e., if the integrated optical depth $\tau \gg 1$. If the atmosphere becomes optically thin, i.e., if $\tau \lesssim 1$, the core can cool on time-scales much shorter than when the core's heat is absorbed and re-radiated by the atmosphere. Therefore, we expect the atmospheric mass that a planet has at the time when the atmosphere becomes optically thin to be retained. 
The optical depth of the whole atmosphere is given by integrating the density profile
\begin{equation}
    \tau = \int_{R_\mathrm{c}}^{R_\mathrm{rcb}} \rho(r) \kappa dr
\end{equation}
which can be evaluated numerically. In the thin limit ($R_\mathrm{rcb} \lesssim 2 R_\mathrm{c}$) and taking $\kappa$ as a constant, the optical depth can be approximated as:
\begin{equation}\label{eq:optdepth}
    \begin{split}
        \tau &\simeq \kappa \rho_\mathrm{rcb} \frac{\gamma-1}{\gamma} R_\mathrm{c}  \bigg(\frac{R_\mathrm{B}'}{R_\mathrm{c}} \bigg)^{1/{\gamma-1}} \\
        &\simeq \bigg(\frac{f}{4.8 \times 10^{-9}}\bigg) \bigg(\frac{\kappa}{0.1\ \mathrm{cm}^2 \mathrm{ g}^{-1}}\bigg) \bigg(\frac{M_\mathrm{c}}{3 M_\oplus}\bigg)^{1/2}
    \end{split} \mathrm{,}
\end{equation}
where we use Equation \ref{eq:Matmthin} to express $\rho_\mathrm{rcb}$ as a function of $f$ and evaluate for $\gamma=7/5$ in the second equality. This approximation differs from the exact integral by less than a factor of two for all relevant atmosphere widths. Equation \ref{eq:optdepth} indicates that the atmosphere transitions to being optically thin at $f \sim 10^{-8}$, equivalent to a surface pressure of a few hundredths of a bar for typical planetary masses. Therefore, we expect core-powered mass loss to cease at residual atmospheric mass fractions of about $10^{-8}$.

\subsection{Combining the different regimes}
\begin{figure}
\centering
\includegraphics[width=\columnwidth]{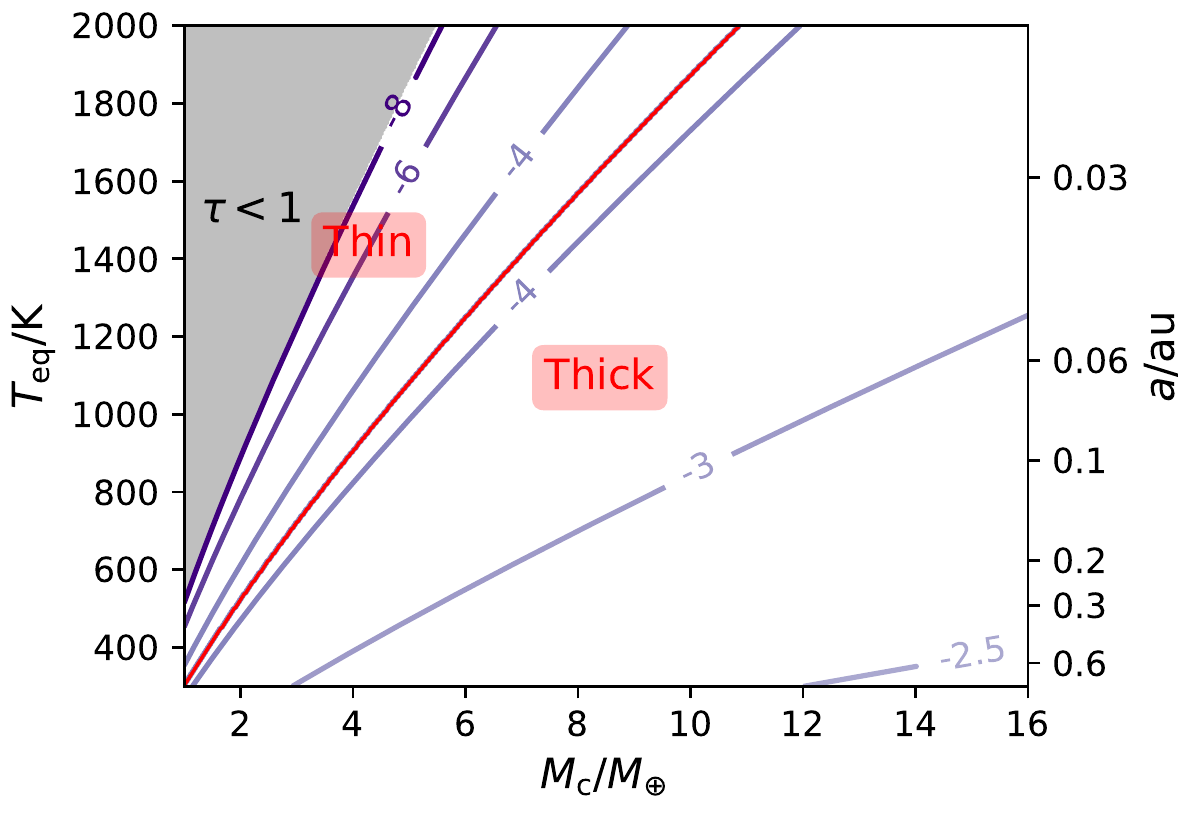}
\caption{Analytical estimates of the final atmospheric mass fraction $f_\mathrm{ret}$ at which $t_\mathrm{cool} = t_\mathrm{loss}$ in the thick and thin regimes taken together. The lines represent contours of $\log_{10}(f_\mathrm{ret})$, ranging from $f_\mathrm{ret} = 10^{-8}$ to $10^{-2.5}$. These $f_\mathrm{ret}$ values are calculated analytically as a function of $M_\mathrm{c}$ and $T_\mathrm{eq}$, using Equation \ref{eq:f_core_thick} in the thick regime and Equation \ref{eq:f_thin} in the thin regime. The analytic thin regime calculations are independent of $f_\mathrm{init}$, so long as planets enter the core cooling regime. The thick regime, as well as the location of the transition between the two, depend on $f_\mathrm{init}$ due to the dependence of $R_\mathrm{rcb, crit}$ on $f_\mathrm{init}$, and the results shown are for $f_\mathrm{init}=0.03$.
The transition between the thick and thin regimes is shown as a red line, and the region of parameter space for which the atmosphere becomes optically thin ($\tau < 1$) are shaded in gray.}
\label{fig:f_contours}
\end{figure}

Now that we have derived the retained atmospheric mass fraction $f_\mathrm{ret}$ in all these regimes of planet evolution, we can combine them into one analytic model. The core can only cool enough to impact the evolution of planets if $E_\mathrm{c} > E_\mathrm{atm}$ at the end of the spontaneous mass loss phase. By equating Equations \ref{eq:energysimple} and \ref{eq:Ecorethin} for $R_\mathrm{rcb} \sim 2R_\mathrm{c}$, we find that a planet's available energy for cooling is mostly stored in the core if the atmospheric mass remaining after spontaneous mass loss $f_\mathrm{sml} \lesssim \mu/\mu_\mathrm{c} \simeq 0.03$. This value corresponds to planets with $f_\mathrm{init} \lesssim 0.15$, since planets typically lose $\sim 80$\% of their initial mass in the spontaneous mass loss phase. We determine whether a planet is in the thick or thin core cooling analytic regime by whether its radiative convective boundary as derived in Section \ref{sec:sml} is greater than or less than $2 R_\mathrm{c}$, respectively. The thick and thin regimes are combined in Figure \ref{fig:f_contours}, showing the final atmospheric masses and surface pressures predicted by this analytic treatment. In this figure we also include the optically thin limit, i.e., the parameter space in which planets reach this limit before $t_\mathrm{cool} < t_\mathrm{loss}$, which is shaded in gray. The thick and thin regimes do not connect smoothly due to their different atmospheric mass and core temperature approximations, but the trends with planet mass and equilibrium temperature continue across the boundary. We expect the smallest and hottest planets to become optically thin before they can cool enough to keep their envelopes. Slightly larger and/or colder planets reach the thin regime and span orders of magnitude in their final envelope masses. Finally, the largest and coldest planets evolve in the thick regime and typically have final envelope mass fractions $f \sim 10^{-3}-10^{-2}$.

\section{Numerical Method and Scaling Comparison}\label{sec:numerical}
To complement the analytical results presented in Section \ref{sec:analytic}, we implement the energy and mass loss evolution described in Section \ref{sec:basics} numerically. In our simulation, we begin with an initial mass fraction, $f_\mathrm{init}$, and radiative-convective boundary location, $R_\mathrm{rcb, init} = R_\mathrm{B}'$. These values, along with the planet's mass and equilibrium temperature, are sufficient initial conditions to calculate the atmospheric profile.

\begin{figure*}
\centering
\includegraphics[width=0.93\textwidth]{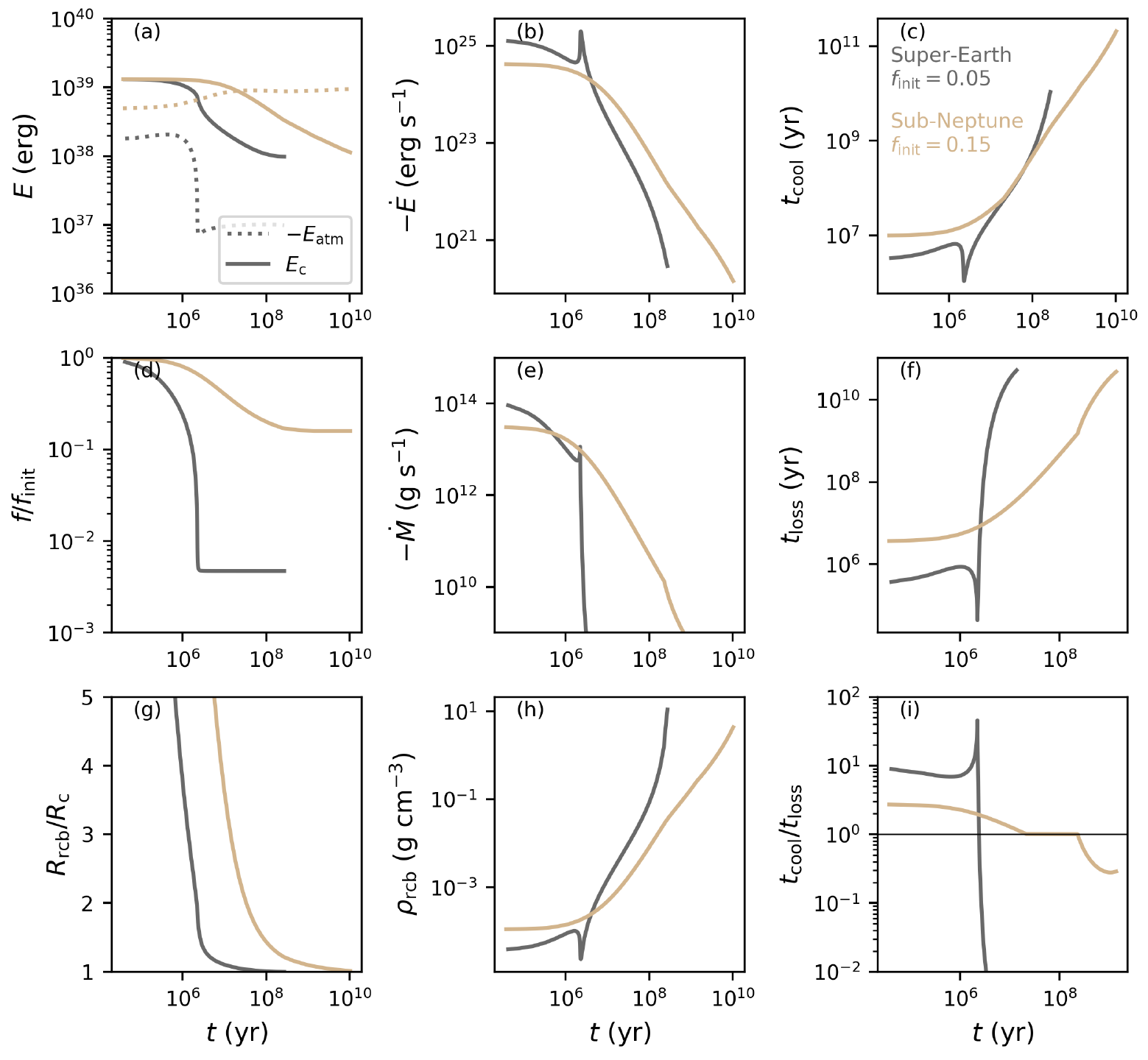}
\caption{Evolution of a super-Earth and a sub-Neptune over time. In the top row we show (a) the total energy of the atmosphere and thermal energy of the core, (b) the energy loss rate, and (c) the cooling time-scale. The middle row displays (d) the atmospheric mass, (e) the mass loss rate, and (f) the mass loss time-scale. Finally on the bottom row we plot (g) the radius of the radiative convective boundary, (h) the density at $R_\mathrm{rcb}$, and (i) the ratio of the cooling and loss time-scales. Both planets have mass $M_\mathrm{c} = 4 M_\oplus$ and equilibrium temperature $T_\mathrm{eq} = 1000\ \mathrm{K}$. The planet shown in gray starts with $f_\mathrm{init}=0.05$, undergoes significant core-powered mass loss, and becomes a super-Earth, but it cools quickly enough to retain a modest H/He envelope: $f_\mathrm{ret} \sim 10^{-4}$. In contrast, the planet shown in tan starts with $f_\mathrm{init}=0.15$, never undergoes core-powered mass loss, and becomes a sub-Neptune, with $f_\mathrm{ret} \sim 3$\%.}
\label{fig:sim_comp_SESN_detailed}
\end{figure*}

We use an Euler method to evolve the atmosphere, calculating the mass and energy lost over a short time-step, $\Delta t = 0.1 \min{(|E_\mathrm{atm}/L|, t_\mathrm{loss})}$, using the mass loss rates and luminosity derived above. We then use the new total mass and energy to derive the new location of the radiative-convective boundary, which results in a new atmospheric profile. This new profile begets new mass- and energy-loss rates, and we carry on evolving the atmosphere. To illustrate the general results of these simulations, we consider in greater detail the super-Earth and sub-Neptune from Figure \ref{fig:sim_comp_SESN_simple}. Figure \ref{fig:sim_comp_SESN_detailed} shows the evolution with time of an expanded set of physical parameters for these two planets. In the top row we show (a) the total energy of the atmosphere and thermal energy of the core, (b) the energy loss rates, and (c) the cooling time-scale. The middle row displays (d) the atmospheric mass, (e) the mass loss rate, and (f) the loss time-scale. Finally on the bottom row we plot (g) the radius of the radiative convective boundary ($R_\mathrm{rcb}$), (h) the density at the $R_\mathrm{rcb}$, and (i) the ratio of the cooling and loss time-scales. As in Figure \ref{fig:sim_comp_SESN_simple}, the planet that becomes a super-Earth, with $f_\mathrm{init} = 0.05$, is shown in gray, and the planet that becomes a sub-Neptune, starting with $f_\mathrm{init} = 0.15$, is in tan.

Initially, mass loss and cooling are rapid in both simulations, and the atmospheres loses mass and contract. In this initial spontaneous mass loss phase, mass loss is limited by the rate at which energy can escape from the planet by diffusion across the radiative-convective boundary. In this regime, the mass-loss rate is directly coupled to the envelope's cooling rate, as shown by comparison of panels (b) and (e) of Figure \ref{fig:sim_comp_SESN_detailed}. In the $f_\mathrm{init} = 0.15$ case shown in tan, the atmosphere's total energy becomes more negative with mass-loss and contraction and starts to exceed in magnitude the core's thermal energy available for cooling, as shown in panel (a). Therefore, the cooling of the core cannot release enough energy to impact the atmospheric evolution significantly. The atmosphere retains $\sim 20$\% of its initial mass at the end of the spontaneous mass loss phase (panel (d)), and it goes on to cool and contract on gigayear time-scales. The evolution of this planet is representative of exoplanets seen in the observed sub-Neptune population.

In contrast, in the $f_\mathrm{init} = 0.05$ case, plotted in gray in Figure \ref{fig:sim_comp_SESN_detailed}, panel (a) shows that the atmospheric energy is much less than the available thermal energy in the core. As the envelope contracts, the temperature of the core decreases. This core cooling releases thermal energy into the atmosphere. In this way, the super-Earth's atmospheric energy increases with time while the core's energy slowly declines. This increase in energy slows the envelope's contraction relative to the sub-Neptune case, as shown by panel (g). While the envelope never stops contracting completely as we approximate in our analytic treatment, this slowing of the contraction sustains mass loss at a faster rate than if the atmosphere were allowed to contract without core heating once the mass-loss rate transitions to being Bondi-limited. The planet thus loses most of its primordial envelope and becomes a super-Earth. This core-powered mass loss phase continues until $\rho_\mathrm{rcb}$ begins to decrease, as shown in panel (h). This decrease makes radiative diffusion through the atmosphere easier, and the envelope's luminosity increases, as demonstrated in panel (b). Therefore, the cooling time-scale, $t_\mathrm{cool}$, decreases (panel (c)). Meanwhile, the mass loss time-scale, $t_\mathrm{loss}$ (panel (f)), increases rapidly once the planet transitions to the Bondi-limited mass-loss regime. In this way, the cooling time-scale becomes less than the mass loss time-scale, shown by the horizontal line in panel (i). Once this occurs, the planet can cool more quickly than mass is lost. As the planet continues to contract, the mass loss rate decreases rapidly due to its exponential dependence on the radiative-convective boundary radius, $R_\mathrm{rcb}$, as shown in panel (e). The mass loss time-scale grows longer than the planet's age (panel (f)), effectively halting loss. This remnant atmosphere then cools and contracts without loss on gigayear time-scales, and the planet becomes a super-Earth with a $f \sim 10^{-4}$ H/He envelope.

\begin{figure}
\centering
\includegraphics[width=\columnwidth]{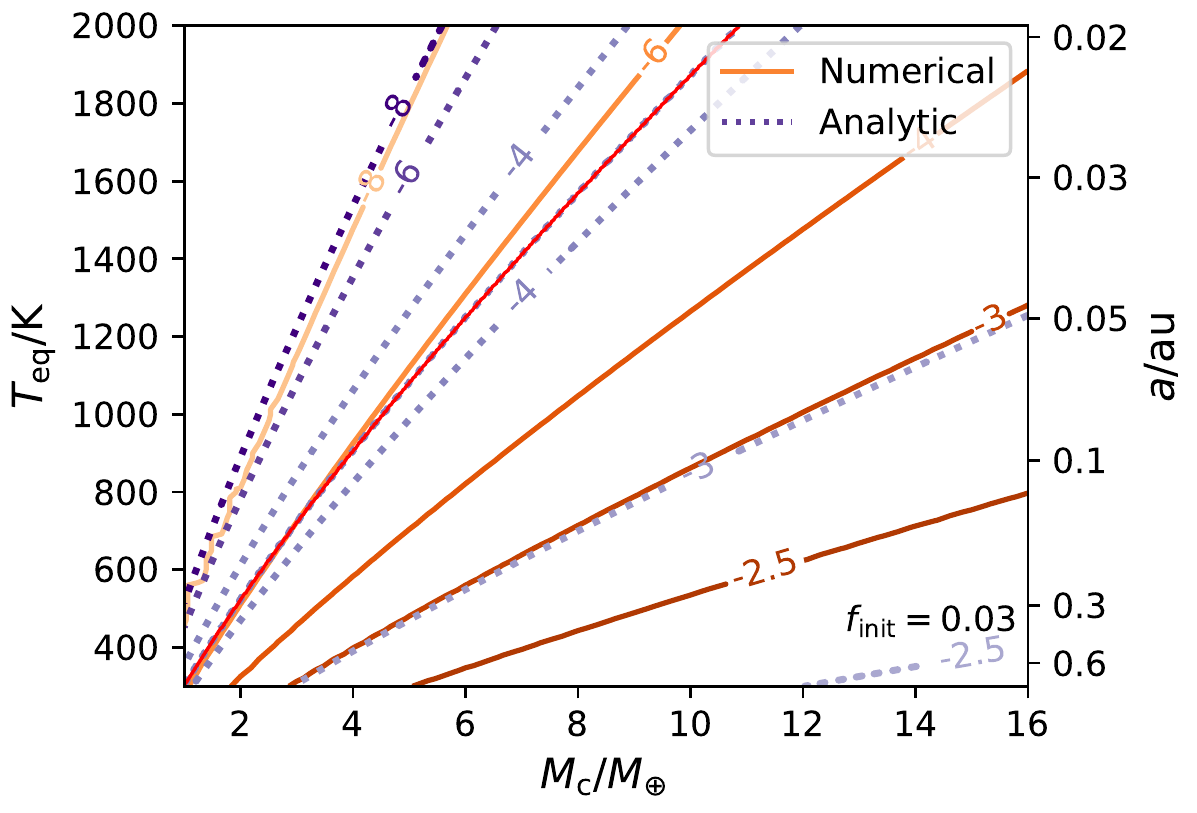}
\caption{Contours of the logarithm of the retained atmospheric mass fractions, $\log_{10} f_\mathrm{ret}$, of an ensemble of $\sim 2500$ planets evolved using our numerical scheme, plotted as functions of core mass, $M_\mathrm{c}$, and equilibrium temperature, $T_\mathrm{eq}$, depicted by orange solid lines. All planets began with $f_\mathrm{init}=0.03$. Overlaid in purple dotted lines are the analytic results previously shown in Figure \ref{fig:f_contours}, including the transition between the thick and thin regimes shown as red line. These results show that our analytic approximations capture the scaling of final atmospheric mass fractions found in our numerical simulations. Differences between the numerical and analytical results include (i) a smooth transition between the thick and thin regimes and (ii) a difference in absolute values of $f_\mathrm{ret}$ due to its dependence on $f_\mathrm{init}$, both of which are not captured in our analytic approximations (see text for details).}
\label{fig:summary_numerical}
\end{figure}

We perform a set of numerical simulations spanning a wide range in parameter space to facilitate comparison between our analytic and numerical results. 
To compare across the entire super-Earth parameter space, Figure \ref{fig:summary_numerical} shows in orange contours of the final atmospheric mass fractions of simulated planets, all starting with the same initial mass fraction, $f_\mathrm{init} = 0.03$, for ease of comparison to our previous analytic results. 
It is apparent from Figure \ref{fig:summary_numerical} that the hotter and less massive planets are able to retain less primordial atmosphere after core-powered mass loss, while cooler and more massive planets can conversely maintain more. These scaling agree quantitatively with our analytic predictions of retained atmospheric mass with core mass and equilibrium temperature (see Figure \ref{fig:f_contours}). These numerical results smooth the discontinuity in our analytic approximations between the thick and thin regimes. We find both numerically and analytically that there exists a  portion of parameter space where super-Earths should be expected to retain modest envelopes of hydrogen and helium. However, the predicted atmospheric mass a planet can retain depends on its initial cooling properties, and hence its initial atmospheric mass fraction, something that is not fully captured in our analytic results.

\begin{figure*}
\centering
\includegraphics[]{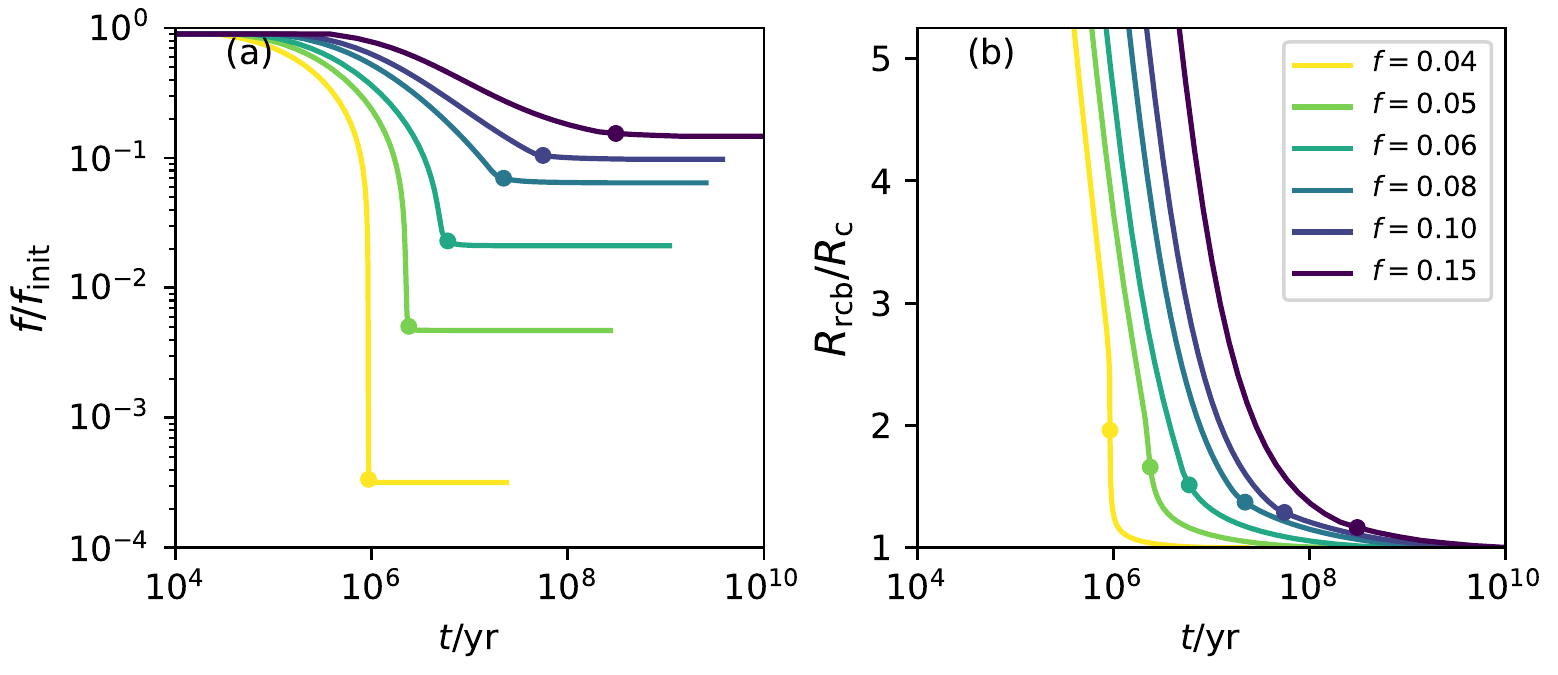}
\caption{Evolution of atmospheric mass, $f$, and radiative-convective boundary radius, $R_\mathrm{rcb}$, over time for seven planets. Each planet has the same core mass, $M_\mathrm{c} = 4 M_\oplus$, and equilibrium temperature, $T_\mathrm{eq} = 1000$ K, but different initial atmospheric mass fractions ranging from 4 to 15 percent. The time at which each planet's cooling time-scale becomes shorter than its mass loss time-scale is denoted by a dot. These planets conclude mass loss with final atmospheric masses varying by many orders of magnitude as a fraction of each planet's initial atmospheric mass. This strong dependence on initial atmospheric mass is due to the different $R_\mathrm{rcb}$ planets are able to contract to as their atmospheres cool and lose mass during the spontaneous mass loss phase.}
\label{fig:different_finit}
\end{figure*}

We show this dependence on initial atmospheric mass in Figure \ref{fig:different_finit}. This figure shows the evolution of seven planets with identical core mass and equilibrium temperature, but which vary in their initial mass fractions at the time of disk dispersal. Figure \ref{fig:different_finit} displays a strong dependence of $f_\mathrm{ret}$ on the initial envelope mass fraction. For example, a planet with $f_\mathrm{init} = 0.04$ retains only $10^{-4}$ times its initial envelope, while a planet with $f_\mathrm{init} \simeq 0.1$ retains $\sim 10$\% of its initial envelope. This dependence on the initial envelope mass fraction can be understood as follows. While the analytic results presented in Section \ref{sec:coreE} are independent of $f_\mathrm{init}$ for a given $R_\mathrm{rcb}$, the $R_\mathrm{rcb}$ at which most mass is lost is determined in part by $f_\mathrm{init}$. We predict this dependence of $R_\mathrm{rcb}$ on $f_\mathrm{init}$ in Equation \ref{eq:rcbapprox}. 
This result arises, because planets with larger initial atmospheric fractions have longer initial atmospheric cooling time-scales, $t_\mathrm{cool} \propto f^2$, per Equation \ref{eq:tcool}. Therefore, the Bondi-limited mass loss rate has to decrease further to become less than the luminosity-limited mass loss rate, equivalent to the atmosphere contracting more. In the highest atmospheric masses we show in Figure \ref{fig:different_finit} ($f_\mathrm{init} \gtrsim 0.10$), the initial atmospheric energies are sufficiently high that the core's cooling has negligible input on atmospheric evolution: these planets never undergo core-powered mass loss at all and become sub-Neptunes. 

In summary, although the analytic and numerical scalings for $f_\mathrm{ret}$ shown in \ref{fig:summary_numerical} are general and independent of $f_\mathrm{init}$, the absolute values of $f_\mathrm{ret}$ do depend on  $f_\mathrm{init}$. The values shown on the contours in Figure \ref{fig:summary_numerical} correspond to initial envelope fractions of 3\%. Since we do not know the initial mass fractions of observed super-Earths \textit{a priori}, these results therefore do not predict the atmospheric masses we expect observed super-Earths to have. Figure \ref{fig:summary_numerical} only compares our numerical simulations to analytic results with the same initial conditions. We discuss possible observational signatures in the following section.

\section{Observational Tests}\label{sec:obs}
A fundamental result of this work is that some super-Earths can retain significant H/He envelopes after core-powered mass loss. This finding implies that super-Earths, which form by core-powered mass loss, can have much lower mean molecular weight atmospheres than planets with entirely outgassed secondary atmospheres. This low mean molecular weight could enhance the detectability of super-Earth atmospheres. The scale height of an atmosphere which has cooled and contracted such that $R_\mathrm{rcb} \simeq R_\mathrm{c}$ is $H = k_\mathrm{B} T_\mathrm{eq}/(\mu g)$, where $g=G M_\mathrm{c}/R_\mathrm{c}^2$ is the gravitational acceleration at the planet's surface. The scale height of a hydrogen-dominated atmosphere, with $\mu=2.2$ amu, is an order of magnitude larger than one composed of even the lightest proposed secondary species, such as water ($\mu=18$ amu). Other possible constituents, such as N$_2$, CO and CO$_2$ have even larger mean molecular weights, and thus atmospheres including them would possess even smaller scale heights. Input of heavier secondary components from surface outgassing would increase the mean molecular weight of any residual primordial atmosphere and therefore decrease its scale height. However, the presence of H/He would increase the atmospheric scale height above what it would be for a purely out-gassed atmosphere.

This large difference in mean molecular weight between residual primordial and pure secondary atmospheres means that super-Earths which retain some primordial hydrogen gas will be distinguishable from those with entirely secondary atmospheres, even if their bulk densities are consistent with a rocky core. Spectroscopic measurements can distinguish between high- and low-$\mu$ atmospheres by the prominence of molecular features, which increases with scale height \citep[e.g.][]{BS12, FM13}. Such measurements have already been performed for sub-Neptunes with the \textit{Hubble Space Telescope} \citep[e.g.][]{BK19,BW19}, and measurements of the hydrogen content of super-Earth atmospheres will be within the capabilities of the upcoming \textit{James Webb Space Telescope} \citep[e.g.][]{GL16,W20} and \textit{Ariel} missions \citep[e.g.][]{E19}. While clouds can obscure these molecular features \citep[e.g.][]{KB14}, high-resolution spectroscopy may enable differentiation even of cloudy high-$\mu$ and low-$\mu$ atmospheres \citep[e.g.][]{GBW20}. These observations will test our hypothesis that some super-Earths can retain primordial H/He, provided that they are not subsequently eroded by photo-evaporation (see discussion in Section \ref{sec:PE} for details). Any residual H/He will also enhance the detectability of other features of interest, such as potential biosignatures \citep[e.g.][]{W20}.

\begin{figure*}
\centering
\includegraphics[width=0.93\textwidth]{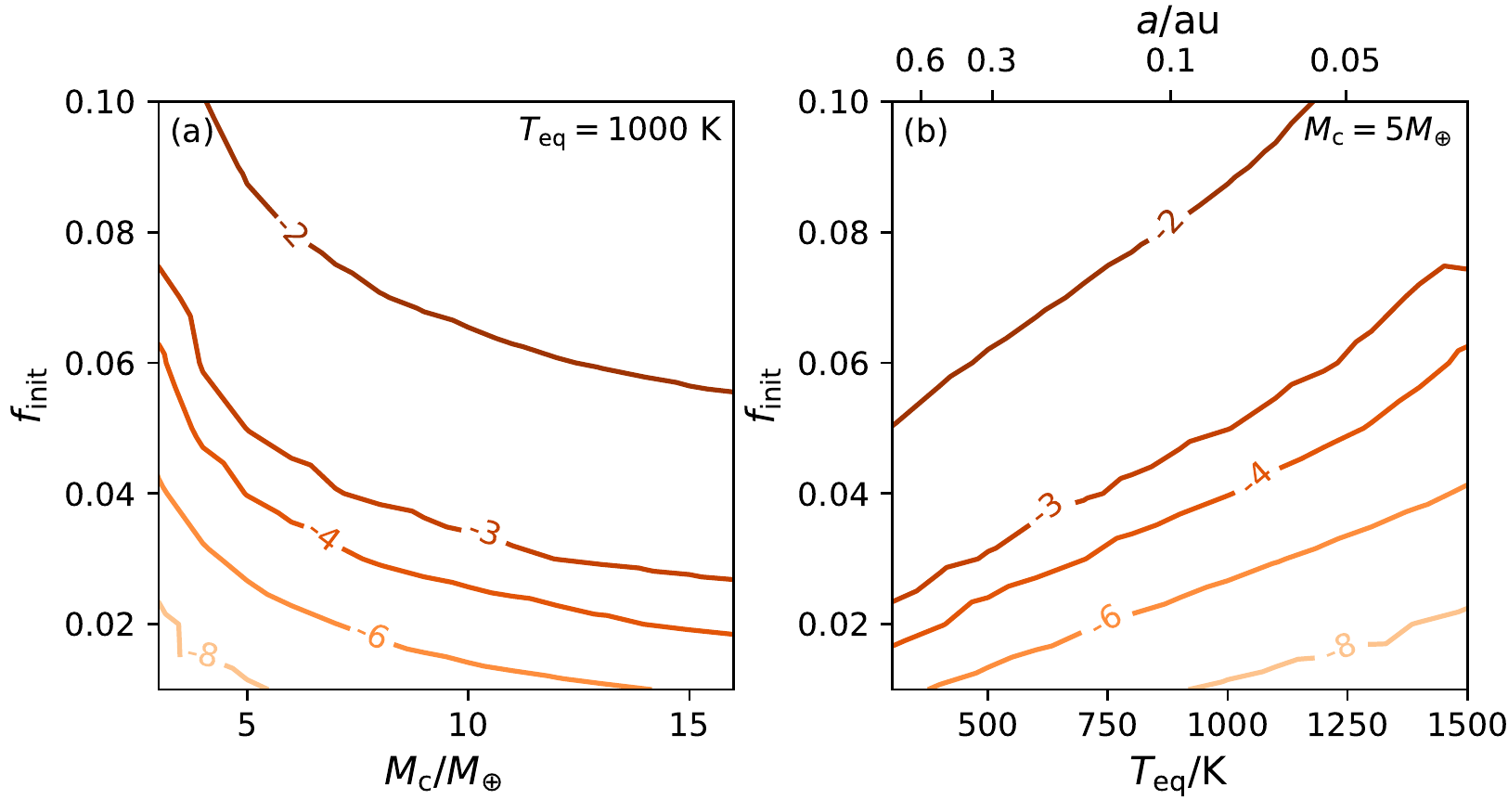}
\caption{Contours of the logarithm of the retained atmospheric mass fraction, $\log_{10}f_\mathrm{ret}$, of an ensemble of numerically-evolved planets, demonstrating the dependence of final atmospheric mass on initial atmospheric mass. In panel (a), each planet has the same equilibrium temperature, $T_\mathrm{eq} = 1000$ K, but a different core mass. In panel (b), each planet instead has the same core mass, $M_c = 5 M_\oplus$, but varies in equilibrium temperature.}
\label{fig:finit_grid}
\end{figure*}

\begin{figure}
\centering
\includegraphics[width=\columnwidth]{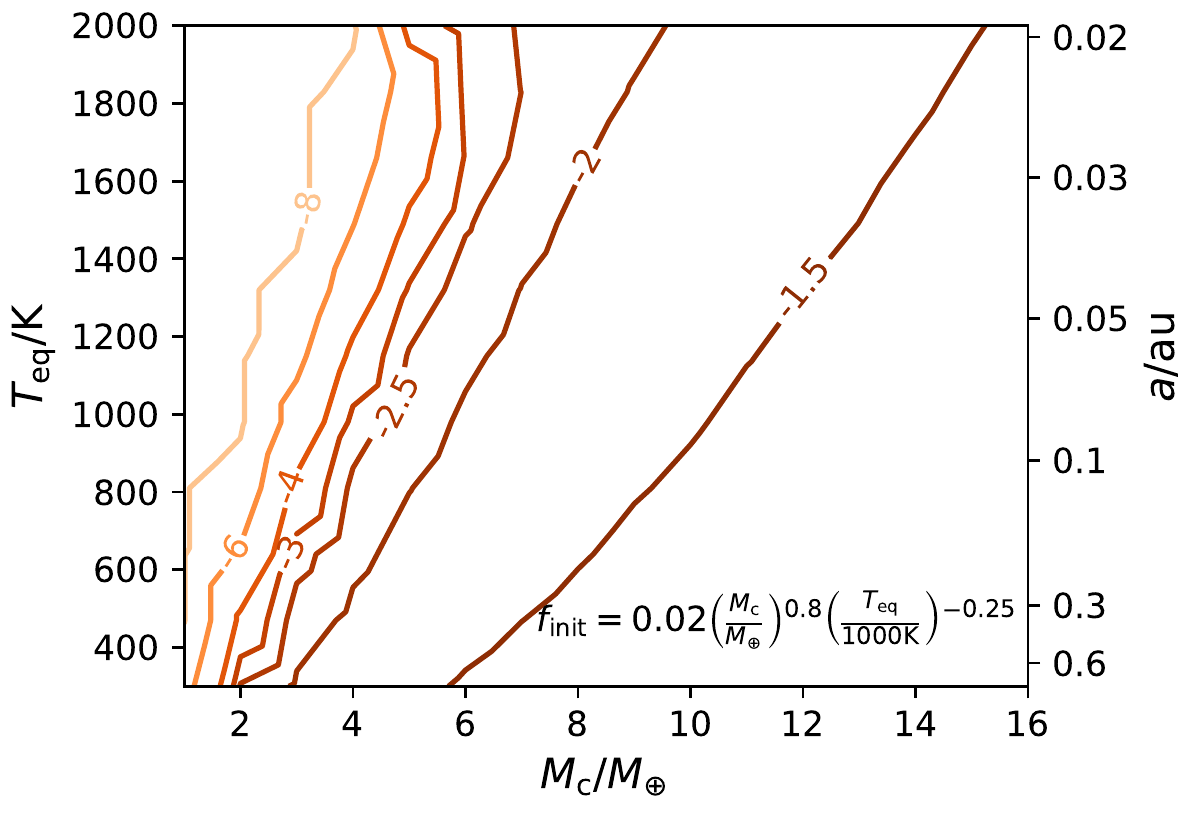}
\caption{Logarithmic contours of the retained atmospheric mass fractions, $\log_{10} f_\mathrm{ret}$, of an ensemble of numerically evolved planets, plotted as a function of core mass, $M_\mathrm{c}$, and equilibrium temperature, $T_\mathrm{eq}$. These planets differ from Figure \ref{fig:summary_numerical} in their initial envelope mass fraction. Instead of assuming $f_\mathrm{init} = 0.03$ for all planets, these planets start with atmospheric masses that scale with core mass and equilibrium temperature as given by $f_\mathrm{init} = 0.02 (M_\mathrm{c}/M_{\oplus})^{0.8} (T_\mathrm{eq}/1000\mathrm{\ K})^{-0.25}$. This scaling is a prediction taken from \citet{GSS16} for a constant disk lifetime of 1 Myr. The transition between sub-Neptune and super-Earth planets is more sudden in this figure than in Figure \ref{fig:summary_numerical}, and smaller mass planets are able to retain H/He envelopes at cooler equilibrium temperatures. Both of these feature are consistent with the observed radius valley \citep{FP17} and past works that demonstrated that the radius valley can be formed by the core-powered mass loss mechanism \citep[e.g.][]{GSS18, GS19}.}
\label{fig:ffunction}
\end{figure}

To test this mechanism of retaining H/He atmospheres, we can predict what atmospheres we would expect planets to retain as a function of their initial atmospheric mass fractions. The initial mass fraction at the time of disk dispersal, $f_\mathrm{init}$, is a major uncertainty which depends on both the planet's cooling time-scale and the disk lifetime. While predictions of the initial atmospheric mass's scaling with parameters such as planet mass, equilibrium temperature, and disk lifetime exist \citep[e.g.][]{GSS16}, their dependence on unknowns such as the disk lifetime make these relations uncertain for observed planetary systems. We therefore cannot predict a unique relationship between a planet's mass and temperature and its final atmospheric mass fraction. However, we can invert the problem and use a super-Earth's residual hydrogen envelope (when observed) to infer its initial atmospheric mass at the time of disk dispersal. To this end, we show in Figure \ref{fig:finit_grid} contours of constant final atmospheric mass fraction, $f_\mathrm{ret}$, as a function of initial atmospheric mass, $f_\mathrm{init}$, and core mass in panel (a). Panel (b) is the same as (a) but the dependence on equilibrium temperature is shown instead of core-mass. 

Furthermore, by using predictions from gas-accretion models for the atmospheric mass fractions as a function of core mass and equilibrium temperature at the time of disk dispersal, we can calculate the expected final envelope masses fractions. For example, in Figure \ref{fig:ffunction} we show the resulting final retained atmospheric mass fractions as a function of core mass and equilibrium temperature using the prediction from \citet{GSS16} given in their Eq. 18, which yield $f_\mathrm{init} = 0.02 (M_\mathrm{c}/M_{\oplus})^{0.8} (T_\mathrm{eq}/1000\mathrm{\ K})^{-0.25}$ for a disk lifetime of 1 Myr.

The transition from massive retained envelopes to nearly-stripped cores with tenuous retained envelopes is more sudden in Figure \ref{fig:ffunction} than in Figure \ref{fig:summary_numerical}, and smaller mass planets are able to retain H/He envelopes if they are at cooler equilibrium temperatures. 
Both of these feature are consistent with the observed radius valley \citep{FP17} and past works that demonstrated that the radius valley can be formed by the core-powered mass loss mechanism \citep[e.g.][]{GS19, GS20}.

\section{Discussion}\label{sec:discussion}
The work presented here demonstrates that some super-Earths can retain small residual H/He envelopes at the end of the core-powered mass loss phase. Such residual H/He envelopes can potentially alter the long-term surface chemistry of these planets and can also increase the scale height of the atmospheres, making atmospheric observations of such planets feasible in the near future. In this section, we discuss the implications of planets having different atmospheric properties than we assume throughout this manuscript. We also quantify the effects of different atmospheric loss processes and discuss how these mechanisms relate to our results.

\subsection{Adiabatic index}
The adiabatic index depends on the number of degrees of freedom of the main constituent of the atmosphere. We assume throughout this work that the atmosphere is primarily diatomic hydrogen. This molecular composition results in an adiabatic index $\gamma=7/5$. However, other values of $\gamma$ are possible due to, for example, the dissociation of hydrogen at high temperatures in the deep interior or heating of the outer parts of the envelope due to absorption of high energy EUV radiation. We thus present our results in terms of $\gamma$ in all analytic expressions throughout the manuscript. Increasing the adiabatic index leads to the mass and energy of the atmosphere being more concentrated in the outer portion of the convective region, leading to more mass loss and less contraction in the spontaneous mass loss phase. However, we do not expect this to significantly alter the overall results of this paper.

\subsection{Opacity}
The opacity at the radiative-convective boundary controls the atmosphere's cooling rate, $L \propto 1/\kappa$, and as such is key to understanding the evolution of super-Earth atmospheres. As discussed in Section \ref{sec:basics}, we use a constant value of the atmosphere's opacity, $\kappa = 0.1$ cm$^2$g$^{-1}$, throughout this work. This choice is based on the typical order of magnitude value expected for H/He-dominated atmospheres \citep[e.g.][]{FML08}. In reality, the atmospheric opacity depends on many factors which are likely to vary over the course of super-Earth atmospheric evolution. More complex opacity models can include a power law increase in opacity with density \citep[e.g.][]{GS19}, and power law increases with temperature and atmospheric metallicity in addition to density \citep[e.g.][]{LC15}. We found that implementing an opacity dependence
on density did not have a major effect on our final atmospheric masses. The strongest effect, on the critical radiative convective boundary to which a planet cools during spontaneous mass loss, is only logarithmic (Equation \ref{eq:rcbapprox}). However, if the overall composition of the atmosphere changes significantly as the super-Earths lose atmospheric mass, an increase in opacity with metallicity over time could affect our results. To test whether we expect the hydrodynamic outflow to significantly fractionate the atmosphere, thereby increasing the metallicity of the remnant atmosphere, we compute the cross-over mass, $\mu_\mathrm{c} = \mu + k_\mathrm{B} T_\mathrm{eq} \dot{M}/(4 \pi b \mu G M_\mathrm{c})$, where $b$ is the binary diffusion coefficient between the two species. This is the molecular weight above which a species cannot be liberated from the planet by the hydrodynamic wind, as the drag forces cannot overcome gravity \citep{H87}. Using $b \sim 5 \times 10^{17} T^{0.75}$ cm$^{-1}$s$^{-1}$, a value typical of the binary diffusion coefficients of common secondary species in hydrogen gas \citep{Z86}, we find – given our mass-loss rates – the smallest value for the cross-over mass for the planets considered in this paper is about $\mu_\mathrm{c} \sim 10^3$ amu. This is much larger than any potential heavy species in these atmospheres. We therefore conclude that secondary species are efficiently carried away by the outflow and that the atmosphere does not become significantly enhanced in heavy elements due to the wind itself for the planets considered in this paper.

Other processes could also influence the atmosphere's opacity as mass loss proceeds. For example, as the overlying pressure decreases, the rate of outgassing may increase \citep[e.g.][]{KB20}, increasing both the metallicity and availability of potential condensates. In this way, the opacity of the atmosphere could increase as primordial H/He is lost. This increase in opacity would decrease the luminosity of the planet and thus increase the cooling time-scale, allowing more mass to be lost before the planet is able to cool on a shorter time-scale than mass is lost. However, this effect may be offset by the increasing difficulty of unbinding the now-heavier atmosphere, as $f_\mathrm{ret} \propto \mu^{7/2} \exp{[\mu]}$, aiding in the retention of whatever mixture of primordial and secondary gas remains. The overall effect of these processes is therefore difficult to quantify without a detailed outgassing and opacity model, which is beyond the scope of this work.

\subsection{Collisionless limit}
The hydrodynamic mass loss rate (Equation \ref{eq:bondiloss}) applies only in the hydrodynamic limit, where a Parker-type outflow is sustained across the sonic point $R_\mathrm{s}$. If the flow becomes collisionless before reaching the sonic point, then the sonic point is no longer causally connected to the radiative-convective boundary and a hydrodynamic out-flow is no longer possible. Instead, in this regime, the loss occurs ballistically via Jeans escape, and is much slower than in the hydrodynamic case \citep[e.g.][]{OM16}.

We can solve for the $R_\mathrm{rcb}$ at which the flow becomes collisionless at the sonic point with the same parameterization as used in Equation \ref{eq:rcbapprox}, again with $\gamma=7/5$:
\begin{equation}\label{eq:R_kn_approx}
\begin{split}
    R_\mathrm{rcb, coll} &= \frac{-R_\mathrm{B}}{\ln{[\epsilon^n R_\mathrm{B}^n/Y]}} \\
    \frac{R_\mathrm{rcb, coll}}{R_\mathrm{c}} &\simeq \cfrac{38.0 \times \bigg(\cfrac{M_\mathrm{c}} {3 M_\oplus}\bigg)^{3/4} \bigg(\cfrac{T_\mathrm{eq}} {1000 \mathrm{\ K}}\bigg)^{-1}}{41.0 -0.5\ln{\bigg[\cfrac{\epsilon}{0.03}\bigg]} +\ln{\bigg[\cfrac{f} {0.05}\bigg]} +2\ln{\bigg[\cfrac{T_\mathrm{eq}} {1000 \mathrm{\ K}}\bigg]} -\ln{\bigg[\cfrac{M_\mathrm{c}} {3 M_\oplus}\bigg]}} \mathrm{,}
\end{split}
\end{equation}
where $Y \equiv \sigma_\mathrm{coll} R_\mathrm{s} \mathrm{e}^2 f_\mathrm{init} M_\mathrm{c}/(8 \pi A \mu R_\mathrm{B}'^{1/(\gamma-1)})$ and in the second line we have taken $\gamma=7/5$. From comparison of the denominators of Equations \ref{eq:rcbapprox} and \ref{eq:R_kn_approx}, the critical radius for collisionless flow is less than the critical cooling radius for all $M_\mathrm{c} > 0.06 M_\oplus$. This implies that cooling will halt mass loss before the outflow becomes collisionless for all the planets we consider here.

\subsection{Stability to photo-evaporation}\label{sec:PE}
In order for these remnant primordial atmospheres to be observed today, they must be stable on gigayear time-scales to a range of processes which could destroy them. One such process is photo-evaporation: mass loss driven by the absorption of high energy radiation from the planet's host star. We can quantify an atmosphere's stability to photo-evaporation after core-powered mass loss by comparing the remaining atmospheric binding energy to the energy the planet receives. The atmospheric energy remaining is $E_\mathrm{atm} \simeq \gamma/ (2\gamma -1) G f M_\mathrm{c}^2 \Delta R_\mathrm{a} /R_\mathrm{c}^2$, where $f$ is the final atmospheric mass fraction \citep{GSS16}. The energy received is the time-integrated high energy flux of the star absorbed by the planet: $E_\mathrm{rec} = E_\mathrm{out} R_\mathrm{rcb}^2/(4 a^2)$, where $a$ is the semi-major axis and we assume the planet's cross-sectional radius is approximately $R_\mathrm{rcb}$. However, photo-evaporation is not perfectly efficient in converting incident energy to mass loss. We quantify the energy available for driving loss as the energy received multiplied by an efficiency factor, $\eta$: $E_\mathrm{av} = \eta E_\mathrm{rec}$. We adopt the nominal value $\eta_\mathrm{nom} = 0.1$, which is based on detailed hydrodynamic simulations \citep[e.g.][]{OW17}. The energy output by the host star, $E_\mathrm{out}$, is given by its time-integrated XUV luminosity. For Sun-like hosts, one commonly-used XUV luminosity evolution track \citep[e.g.][]{OW17} is that found by \citet{JDW12}: $L_\mathrm{XUV} = 1.2 \times 10^{33} (t/\mathrm{Myr})^{-1.5}$ erg s$^{-1}$ for $t>100$ Myr. For this nominal evolution track, the integrated energy output from 100 Myr to 1 Gyr is $E_\mathrm{out,nom} = 5.2 \times 10^{45}$ erg.

Comparing these two energies gives us a criterion for determining the atmosphere's stability to photo-evaporation, $\Phi \equiv E_\mathrm{atm}/E_\mathrm{av}$. Expressing this criterion in terms of planet parameters, we find:

\begin{equation}\label{eq:PE}
\begin{split}
    \Phi &\simeq \frac{\gamma}{2\gamma-1} \frac{G f M_c^2 a^2}{R_\mathrm{rcb}^2 R_\mathrm{c} \eta E_\mathrm{out}}\\
    &\simeq \!\begin{multlined}[t] \frac{f}{3.3 \times 10^{-3}} \bigg(\frac{E_\mathrm{out}} {5.2 \times 10^{45} \mathrm{erg}}\bigg)^{-1} \bigg(\frac{\eta} {0.1}\bigg)^{-1} \\
        \bigg(\frac{M_\mathrm{c}}  {3M_\oplus}\bigg)^{5/4} \bigg(\frac{T_\mathrm{eq}} {1000 \mathrm{K}}\bigg)^{-4} \bigg(\frac{R_\mathrm{rcb}} {R_\mathrm{c}}\bigg)^{-2} \mathrm{.}
    \end{multlined}
\end{split}
\end{equation}
If $\Phi \lesssim 1$, then sufficient energy is received to photo-evaporate the atmosphere. It is clear from Equation \ref{eq:PE} that many of the tenuous atmospheres we predict are vulnerable to photo-evaporation on gigayear timescales. However, the order of magnitude observed difference in the XUV luminosity evolution of fast- and slow-rotating stars of the same type \citep{T15} can have major impacts on the gigayear-time-scale evolution of super-Earths and sub-Neptunes \citep[e.g.][]{P21}. Specifically, a planet's remnant atmosphere is more resistant to erosion by photo-evaporation if the planet's host star emits less XUV energy than implied by the XUV evolution track of \citet{JDW12}, and/or if the photo-evaporative efficiency is lower: $\Phi \propto (E_\mathrm{out} \eta)^{-1}$.

\begin{figure}
\centering
\includegraphics[width=\columnwidth]{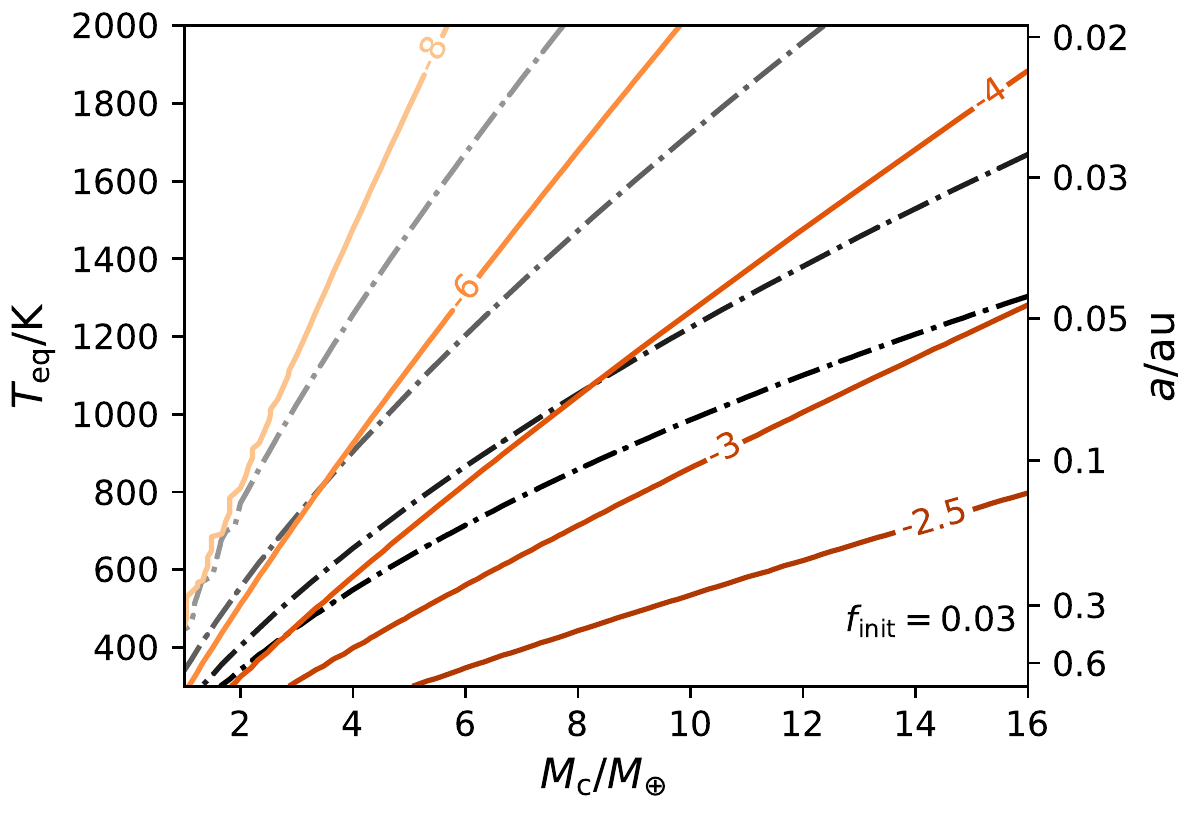}
\caption{Contours of the atmospheric stability criterion, $\Phi=1$ (see Equation 34), for the ensemble of $\sim 2500$ simulated planets with $f_\mathrm{init} = 0.03$ previously presented in Figure \ref{fig:summary_numerical}. The $\Phi=1$ contours are plotted in gray dash-dotted lines. The shades represent variations in the product of the XUV energy output by the host star and the photo-evaporative efficiency, scaled to our adopted nominal values: $E_\mathrm{out} \eta/(E_\mathrm{out,nom} \eta_\mathrm{nom})=10^{-5}$, $10^{-3}$, $10^{-1}$, and $1$ from lightest to darkest. The residual atmospheric mass fractions, $\log_{10} f_\mathrm{ret}$, at the end of core-powered mass-loss are shown as solid orange contours and are identical to those of Figure \ref{fig:summary_numerical}. The residual atmosphere of a planet lying along a particular contour (shown in solid orange) is stable to photo-evaporation if it falls below a given photo-evaporation contour.}
\label{fig:PE_effect}
\end{figure}

We examine this vulnerability in more detail in Figure \ref{fig:PE_effect}. In this figure, we plot the stability criterion, $\Phi = 1$, in gray dash-dotted lines for the residual atmospheres of the suite of simulated planets with $f_\mathrm{init} = 0.03$ previously presented in Figure \ref{fig:summary_numerical}. The different shades represent different values of the product of the XUV energy output by the host star and the efficiency factor, scaled to the nominal values: $E_\mathrm{out} \eta/(E_\mathrm{out,nom} \eta_\mathrm{nom})=10^{-5}$, $10^{-3}$, $10^{-1}$, and $1$ from lightest to darkest. We take $E_\mathrm{atm}$ as the atmosphere's energy after it cools and contracts to $R_\mathrm{rcb} \sim R_\mathrm{c}$. Meanwhile, $E_\mathrm{av}$ is the energy the planet would receive between 100 Myr and 1 Gyr of its host star's evolution. These calculations are overlain on the $f_\mathrm{ret}$ values previously depicted in Figure \ref{fig:summary_numerical}. The residual atmosphere of a planet lying along a particular contour (shown in solid orange) is stable to photo-evaporation if it falls below a given photo-evaporation contour. Figure \ref{fig:PE_effect} demonstrates that only the residual atmospheres of cooler and more massive planets can survive photo-evaporation, unless a planet orbits a slow rotator with low XUV output. Smaller residual atmospheres could also be preserved if photo-evaporation is much less efficient than typically assumed. 
In this way, observations of super-Earth atmospheres provide a test of the efficacy of photo-evaporation. If these primordial atmospheres can survive to be observed, then photo-evaporation may not be as effective on long time-scales as predicted.

The argument presented above only accounts for the effects of photo-evaporation after the first 100~Myr once the XUV luminosity has decayed from its peak intensity. However, it is possible, in some regions of parameter space, for super-Earths to form by core-powered mass loss on time-scales shorter than this. In this case, the energy received from XUV radiation would be significantly increased. To determine the final envelope mass fraction in this case requires a coupled core-powered mass loss and photo-evaporation model. Such a combined model is beyond the scope of this paper but is planned for future work.

\section{Summary \& Conclusions}\label{sec:conclusions}
In this work, we have demonstrated that despite appearing to be bare cores from their radii and bulk densities, some super-Earth planets may possess thin primordial H/He envelopes at the end of the core-powered mass loss phase. This occurs because after rapid hydrodynamic loss strips a super-Earth of the bulk of its atmosphere, the accompanying decrease in density at the radiative-convective boundary allows the envelope to cool more efficiently. Once the cooling time-scale becomes shorter than the mass loss time-scale, the envelope can contract, which preserves the remaining primordial gas. This occurs even if sufficient heat energy remains in the core to unbind the entire atmosphere, as its cooling is mediated by radiative transfer through the envelope. Super-Earths with larger core masses and lower equilibrium temperatures retain larger mass fractions of gas. These fractions range from negligible ($f \sim 10^{-8}$) to much thicker than the present-day atmosphere of Venus ($f > 10^{-4}$). These retained envelopes following core-powered mass loss differ from the predicted outcomes of primordial super-Earth atmospheres in photo-evaporation models. Under photo-evaporative mass loss, as the atmosphere loses mass, the high-energy photons penetrate further into the atmosphere, and heating and loss become more effective. Thus if sufficient energy is received, the expectation is that photo-evaporation completely unbinds the original envelopes, unlike core-powered mass loss. However, many of the tenuous primordial atmospheres, which we predict core cooling to save, are susceptible to long-term erosion by photo-evaporation, at least for nominal photo-evaporation models. Higher mass planets further from their stars are less vulnerable to photo-evaporative stripping, as are those orbiting slower rotators with lower XUV outputs.

Any residual H/He atmosphere could significantly affect redox states and the resultant chemistry on the surfaces of these early planets, as well as the initial conditions for outgassing. Any H/He dominated super-Earth atmospheres would have mean molecular weights close to those of the primordial disk, $\mu \sim 2.2$ amu. Such a composition would lead to scale heights an order of magnitude larger than would be expected from purely secondary atmospheres, which are usually expected to be made of heavier constituents. A lower mean molecular weight would make these super-Earths more favorable for spectroscopic observations, and H-rich super-Earths should be distinguishable from those with pure secondary atmospheres by their scale heights in the near future.

\section*{Acknowledgements}
We thank James Owen and Sivan Ginzburg for insightful comments which improved the manuscript. H.E.S. gratefully acknowledges support from the National Aeronautics and Space Administration under grant No. 
 $17~\rm{NAI}18\_~2-0029$ issued through the NExSS Program. In this work we use the \textsc{numpy} \citep{numpy}, \textsc{matplotlib} \citep{Matplotlib}, and \textsc{scipy} \citep{scipy} packages.

\section*{Data Availability}
Data available on request.



\bibliographystyle{mnras}
\bibliography{references} 


\bsp	
\label{lastpage}
\end{document}